\documentclass[
aps,
prd,
12pt,
nopreprintnumbers,
showpacs,
eqsecnum,
nofootinbib
]{revtex4-1}

\usepackage{graphicx}
\usepackage{amssymb}

\begin{document}

\title{Classical and quantum cosmology of $K$-essentially modified $R^2$ and
pure $R^p$ gravity}
\author{Nahomi Kan}\email[]{kan@gifu-nct.ac.jp}
\affiliation{National Institute of Technology, Gifu College,
Motosu-shi, Gifu 501-0495, Japan}
%\author{Masashi Kuniyasu}\email[]{mkuni13@yamaguchi-u.ac.jp}
\author{Kiyoshi Shiraishi}\email[]{shiraish@yamaguchi-u.ac.jp}
%\author{Kohjiroh Takimoto}\email[]{i016vb@yamaguchi-u.ac.jp}
\author{Mai Yashiki}\email[]{g005wb@yamaguchi-u.ac.jp}
\affiliation{
Graduate School of Sciences and Technology for Innovation, Yamaguchi
University, Yamaguchi-shi, Yamaguchi 753--8512, Japan}
\date{\today}
%\date{}

\begin{abstract}
We present a gravitational action with a modified higher order term 
of a combination of scalar curvature and Lagrangian density of a scalar field.
This type of models has been considered first by Cruz-Dombriz {\it et al.}
The classical and quantum cosmologies governed by the modified action
 are studied.
Models described by a positive-definite action and a pure arbitrary-powered
scalar curvature action without the standard Einstein--Hilbert term are also
investigated. We show some particular cases in which exact solutions can be
obtained.

\end{abstract}

%\preprint{}

\pacs{%
%02.10.Ox, %%%Combinatorics; graph theory
%02.20.Sv, %Lie algebra of Lie groups
%02.30.Hq, %Ordinary differential equations
%04.20.-q, %%%Classical general relativity
%04.20.Fy, %%Canonical formalism, Lagrangians, and variational principles
04.20.Jb, %%Exact solutions
%04.25.-g, %Approximation
%04.25.Nx, %%%Post-Newtonian approximation; perturbation theory; related
%approximations
%04.40.-b, %Self-Gravitating systems
%04.40.Nr, %%Einstein-Maxwell spacetime
04.50.-h, %%%%%Higher-dimensional gravity and other theories of gravity 
%04.50.Cd, %Kaluza-Klein theories 
%04.50.Gh, %Higher-dimensional black holes, black strings, 
%and related objects 
04.50.Kd, %%%Modified theories of gravity 
%04.60.-m, %%Quantum gravity
%04.60.Kz, %%Lower dimensional models; minisuperspace models
%04.60.Rt, %Topologically massive gravity
%04.62.+v, %Quantum fields in curved spacetime
%04.65.+e, %Supergravity
%04.70.Bw, %%%Classical black holes
%05.30.Jp, %Boson systems
%11.10.-z, %%%Field theory
%11.10.Lm, %%%Nonlinear or nonlocal theories and models 
%11.10.Kk, %%%Field theories in dimensions other than four
%11.25.-w, %Strings and branes
%11.25.Mj, %%Compactification and four-dimensional models
%11.27.+d %%Extended classical solutions; cosmic strings, 
%domain walls, texture 
%12.60.-i, %Models beyond the standard model
%95.35.+d, %Dark matter
%95.36.+x, %Dark energy
%98.80.-k, %%%Cosmology 
98.80.Cq, %%%%%Particle-theory and field-theory models of the early
%Universe  
%98.80.Dr, %Relativistic cosmology 
98.80.Qc%Quantum cosmology
%98.80.Jk% %%Mathematical and relativistic aspects of cosmology
.}

\maketitle

%%%%%%%%%%%%%%%%%%%%%%%%%%%%%%%%%%%%%%%%%%%%%%%%%%%%%%%%%%%%%%%%%%%%%%%%%%%
%%%%%%%%%%%%%%%%%%%%%%%%%%%%%%%%%%%%%%%%%%%%%%%%%%%%%%%%%%%%%%%%%%%%%%%%%%%
%%%%%%%%%%%%%%%%%%%%%%%%%%%%%%%%%%%%%%%%%%%%%%%%%%%%%%%%%%%%%%%%%%%%%%%%%%%
\section{Introduction}
\label{sec1}
%%%%%%%%%%%%%%%%%%%%%%%%%%%%%%%%%%%%%%%%%%%%%%%%%%%%%%%%%%%%%%%%%%%%%%%%%%%
%%%%%%%%%%%%%%%%%%%%%%%%%%%%%%%%%%%%%%%%%%%%%%%%%%%%%%%%%%%%%%%%%%%%%%%%%%%
%%%%%%%%%%%%%%%%%%%%%%%%%%%%%%%%%%%%%%%%%%%%%%%%%%%%%%%%%%%%%%%%%%%%%%%%%%%

Recent developments of observational cosmology suggest
that our universe experienced two acceleration eras:
the inflation era in the very early times \cite{inflation,EW} and the era of
late-time acceleration in the present time
\cite{darkenergy1,darkenergy2,darkenergy3}.

Inflation, a rapid expansion of the very early universe, is supposed to be caused
by an evolving scalar field (inflaton) coupled to gravity.
Although many mechanisms which bring about inflation have been proposed until now, 
it is believed that most of favorable models use the slow-roll regime,
i.e., the value of the inflaton changes slowly in the cosmic time.

On the other hand, since the general theory of relativity can merely be regarded as
a low-energy theory, it is considerable that the complete gravitational theory
is unknown yet. Therefore, various modifications of Einstein gravity have been
studied by many authors
\cite{NO1,SF,CLF,FT,NO2,CL,CFPS,Koyama,NOO,Ishak}. Among these, the
$R^2$-inflationary model (a.k.a. the Starobinsky model
\cite{Starobinsky,Vilenkin,MMS1}) is not also the earliest model which contains
quantum corrections to the Einstein gravity but an excellent model of
cosmic inflation whose predictions agree with recent observations \cite{EW}.

The $R^2$ gravity (where $R$ means a scalar curvature) is a higher derivative
theory that can be reduced to second order equations through a redefinition of 
variables. In Einstein frame, the model
contains a scalar mode with an almost flat potential. We call this mode a scalaron.
The scalaron plays a role of a slow-roll inflaton in the model and
can explain almost scale-invariant perturbations from 
stretched quantum fluctuations.

In this paper, we explore some possibilities of the theoretical extension of
$R^2$ gravity.
The Starobinsky model with a minimal scalar matter field has been considered by
many authors \cite{BP,BDP,MKW,KK,CCH,MMS,SN}. 
Some authors considered the extension in order to investigate scenarios of seeding
curvature perturbations by the scalar field, while some authors are motivated by
chaotic-type inflationary models. Incidentally, there are other studies on the
models with the
$R^2$ term, which examine a possible improvement of the Higgs inflation
\cite{GT1,BOT,SM,Wang,Ema,Salvio,MSY,GT2,GS}.

Recently, Cruz-Dombriz {\it et al.}\cite{CEOS} proposed various models of gravity
with nonstandard couplings to a scalar field. They considered that
the `$K$-essence' \cite{AMS}, such as a form of kinetic term of a scalar field,
couples with the modified gravity. Our models considered in the present paper are
much akin to one of their models of `non-minimally coupled $K$-essence'
\cite{CEOS}. The present models 
lead to  simple dynamics of an additional scalar field in classical and
quantum cosmologies. The additional scalar can behave as an inflaton or a
quintessence for late-time acceleration.

This paper is organized as follows:
In the next section, we consider the $K$-essential modification of $R^2$ gravity
and discuss its properties.
Quantum cosmology of the model is investigated in Sec.~\ref{qc}. 
Classical and quantum  properties of the model of an extension of pure $R^2$ theory
is studied in Sec.~\ref{pd}. In Sec.~\ref{monomial}, the $K$-essential extension of
pure $R^p$ gravity, where $p$ is an arbitrary number, is studied. Finally, we
conclude the present study in Sec.~\ref{dis}.

In Appendix \ref{AA}, we revisit the comparison in known exact solutions
for pure $R^2$ quantum cosmology.

%%%%%%%%%%%%%%%%%%%%%%%%%%%%%%%%%%%%%%%%%%%%%%%%%%%%%%%%%%%%%%%%%%%%%%%%%%%
%%%%%%%%%%%%%%%%%%%%%%%%%%%%%%%%%%%%%%%%%%%%%%%%%%%%%%%%%%%%%%%%%%%%%%%%%%%
%%%%%%%%%%%%%%%%%%%%%%%%%%%%%%%%%%%%%%%%%%%%%%%%%%%%%%%%%%%%%%%%%%%%%%%%%%%
\section{classical cosmology of the extension of $R^2$ gravity}
\label{cc}
%%%%%%%%%%%%%%%%%%%%%%%%%%%%%%%%%%%%%%%%%%%%%%%%%%%%%%%%%%%%%%%%%%%%%%%%%%%
%%%%%%%%%%%%%%%%%%%%%%%%%%%%%%%%%%%%%%%%%%%%%%%%%%%%%%%%%%%%%%%%%%%%%%%%%%%
%%%%%%%%%%%%%%%%%%%%%%%%%%%%%%%%%%%%%%%%%%%%%%%%%%%%%%%%%%%%%%%%%%%%%%%%%%%

The Starobinsky model is defined by the action \cite{Starobinsky,Vilenkin,MMS1}
\begin{equation}
S_{S}=\int
d^4x\,\sqrt{-g}\,\left[\alpha R+\frac{\beta}{2}R^2\right]
\,,
\end{equation}
where $R$ is the scalar curvature. The coefficients $\alpha$ and $\beta$ are
constants.
Our starting point in the present study is the modified action
\begin{equation}
S_{S}=\int
d^4x\,\sqrt{-g}\,\left\{\alpha
\left[R-\frac{1}{2}(\nabla\phi)^2-
V(\phi)\right]+\frac{\beta}{2}\left[R-\frac{1}{2}(\nabla\phi)^2-
\gamma V(\phi)\right]^2\right\}
\,,
\label{ours}
\end{equation}
where $(\nabla\phi)^2=g^{\mu\nu}\nabla_\mu\phi\nabla_\nu\phi$, and
$V(\phi)$ is the potential of the scalar field $\phi$. 
A difference from the action of Cruz-Dombriz {\it et al.}~\cite{CEOS} is an
introduction of a parameter $\gamma$ in front of the potential $V(\phi)$ in the
square bracket.

The action (\ref{ours}) is classically equivalent to
\begin{equation}
S=\int d^4x\sqrt{-g}\left\{\alpha
\left[R-\frac{1}{2}(\nabla\phi)^2-
V(\phi)\right]+\beta\chi\left[R-\frac{1}{2}(\nabla\phi)^2-
\gamma V(\phi)\right]-\frac{\beta}{2}\chi^2\right\}\,,
\label{ours2}
\end{equation}
since the equation of motion with respect to the auxiliary field $\chi$,
$\frac{\delta{\cal S}}{\delta\chi}=0$, implies 
$\chi=R-\frac{1}{2}(\nabla\phi)^2-
\gamma V(\phi)$.

We can eliminate the $\chi$-dependence in front of the Einstein--Hilbert term $R$
in the action (\ref{ours2}) by a Weyl transformation.
In other words, we consider a Weyl-transformed metric $\tilde{g}_{\mu\nu}$
which satisfies
$\sqrt{-g}(\alpha+\beta\chi)R=\sqrt{-\tilde{g}}\alpha\tilde{R}+\cdots$,
where $\tilde{R}$ is the Ricci scalar constructed from $\tilde{g}_{\mu\nu}$.
To this end, we choose 
$g_{\mu\nu}=\left(1+\frac{\beta}{\alpha}\chi\right)^{-1}\tilde{g}_{\mu\nu}$.
Then, we obtain
\begin{equation}
S=\int d^4x\sqrt{-\tilde{g}}\,\,\alpha\left[
\tilde{R}-\frac{1}{2}(\tilde{\nabla}\phi)^2-
\frac{1}{2}(\tilde{\nabla}\psi)^2-U(\psi,\phi)\right]\,,
\end{equation}
where
\begin{equation}
U(\psi,\phi)=
\frac{1}{2\beta'}\left(e^{\frac{1}{\sqrt{3}}\psi}-1\right)^2+\left[\gamma
e^{\frac{1}{\sqrt{3}}\psi}-(\gamma-1)e^{\frac{2}{\sqrt{3}}\psi}\right]V(\phi)\,.
\label{potg}
\end{equation}
Here, we defined $\beta'\equiv\beta/\alpha$ and introduced the new scalar variable
$\psi\equiv-\sqrt{3}\ln\left(1+{\beta'}\chi\right)$.
The boundary terms in the action have been omitted.

Note that the scalar field $\phi$ has a canonical kinetic term in our
model.  Contrary to this, in the simplest extension of the Starobinsky model
\cite{BP,BDP,MKW,KK,CCH,MMS,SN}, the kinetic term of the additional scalar field
$\phi$ couples to the scalaron field $\psi$ through the exponential function,
such as $e^{\frac{1}{\sqrt{3}}\psi}(\tilde{\nabla}\phi)^2$.
On the other hand, the potential $U(\psi,\phi)$ coincides with the one of 
the simplest extension of the Starobinsky model \cite{BP,BDP,MKW,KK,CCH,MMS,SN}, if
we set $\gamma=0$ in (\ref{potg}).

Cruz-Dombriz {\it et al.}~\cite{CEOS} considered the case with $\gamma=1$, in
parametrization used here. We claim that an interesting choice for the parameter
$\gamma$ is, however, 
$\gamma=2$. 
%R1
One can observe that the trace of the Einstein equation from the lowest order
terms, which is obtained by setting $\beta=0$ in our model, gives
$R-\frac{1}{2}(\nabla\phi)^2-2V(\phi)=\chi=0$.
%R1
Then, $\frac{\partial U}{\partial\psi}$ vanishes when $\psi=0$ for any
values of $\phi$. 
%Because this case is very typical, we consider that $\gamma=2$,
%unless the value of $\gamma$ is particularly mentioned.

One can find that the present model has advantages and disadvantages than the
simplest extension. The non-canonical kinetic term found in
\cite{BP,BDP,MKW,KK,CCH,MMS,SN} induces an additional frictional effect in the
evolution of the scalar field $\phi$. If the slow-roll motion of the scalar field
is required, the canonical kinetic term brings about a demerit. Also,
interesting processes due to the kinematic coupling with the scalaron $\psi$
are known, for reheating process and generation of perturbations in the universe
\cite{BDP,GW,STY,WBMR,MF,LLPT,Wands,BR,WW}. These are disadvantages.

Conversely, the canonical form of the additional scalar field can be said to
make the model simple. 
The advantage of our model with the standard kinetic term is also found in direct
applications of quantum cosmology in the known two-field models. This will be
discussed in the next section.

%R1
Now, we examine the parameter dependence of our present model.
We adopt here $V(\phi)=\frac{\rho}{3\beta'}\phi^2$ as a typical case.
If $\rho=1$, we find the
$\frac{\partial^2U}{\partial\phi^2}=\frac{\partial^2U}{\partial\psi^2}$
at $\psi=\phi=0$.

\begin{figure}[ht]
\centering
\includegraphics[width=5cm]{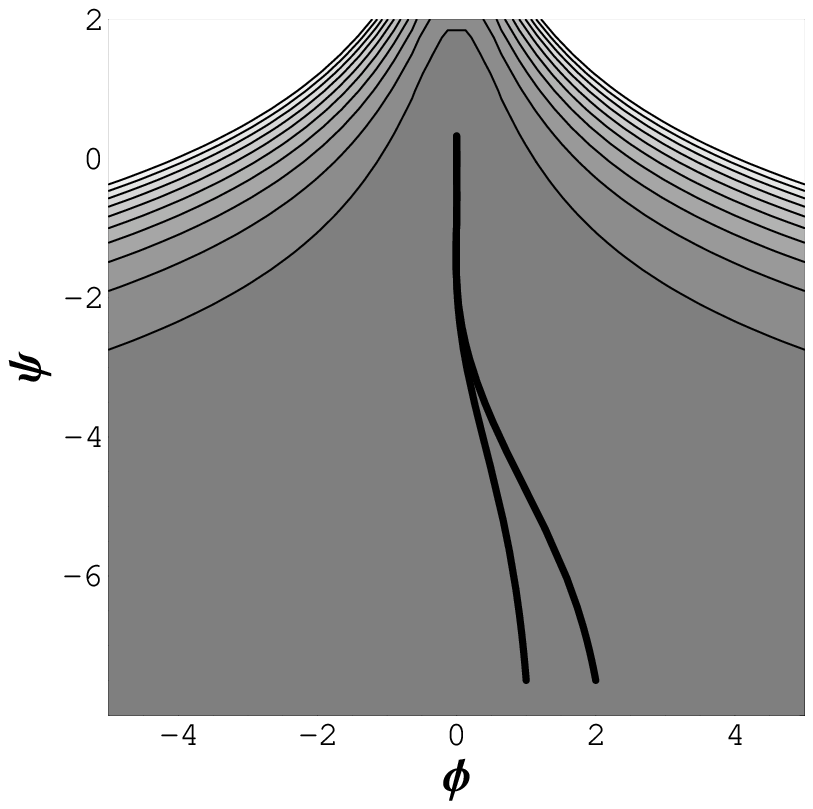}
\includegraphics[width=5cm]{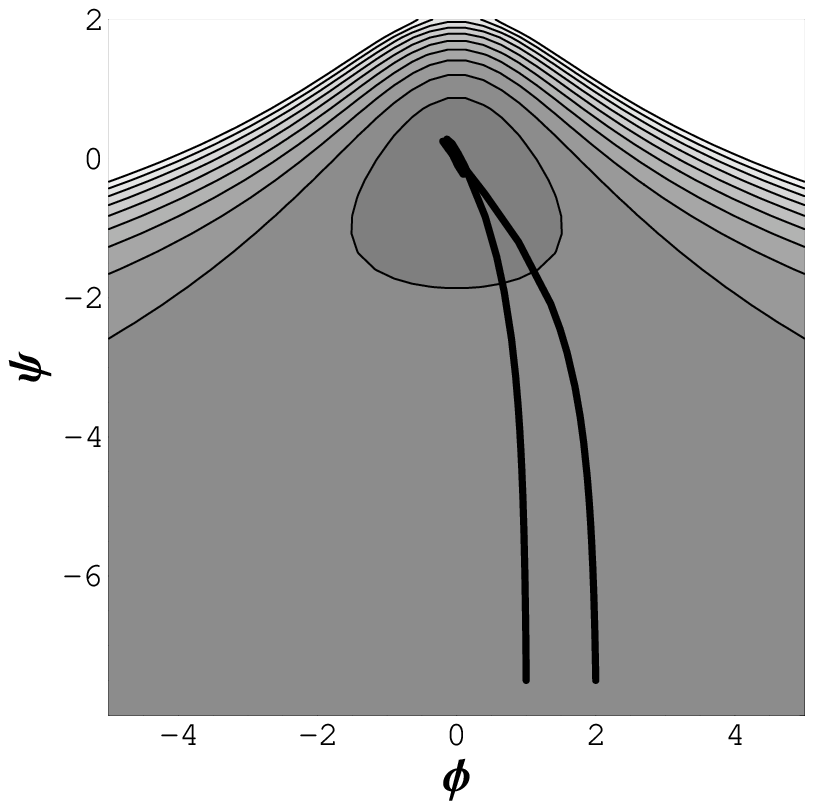}
\includegraphics[width=5cm]{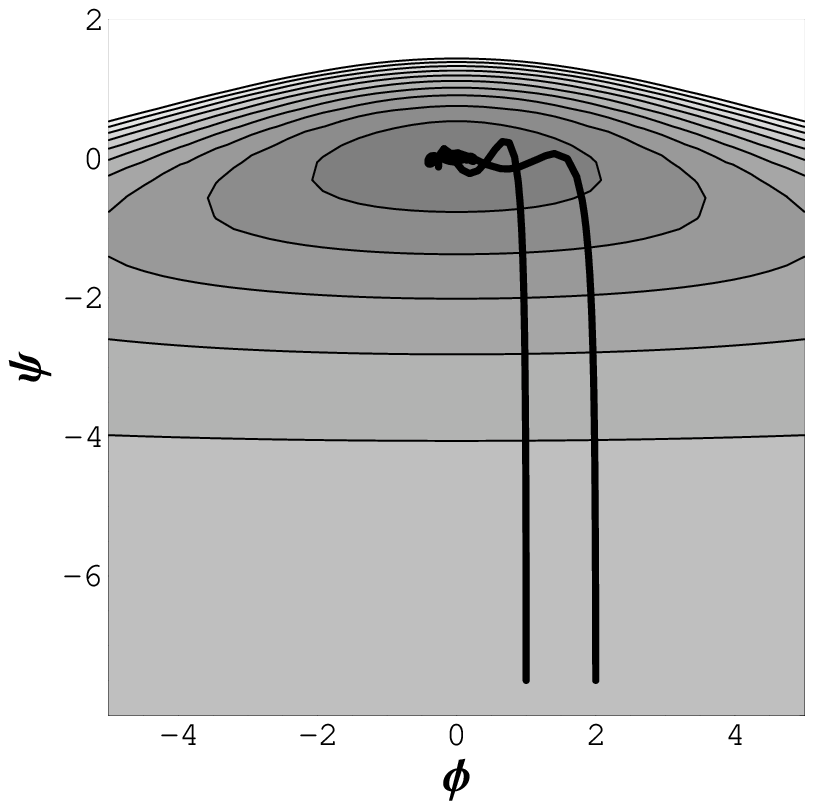}
\\
\hspace{0.5cm} (a) \hspace{4.5cm} (b) \hspace{4.5cm} (c)
\caption{Typical trajectories, on the contour plot of the potential $U(\psi,\phi)$,
for: (a)
$\rho=10$, (b)
$\rho=1$, (c)
$\rho=0.1$, in the case with $\gamma=0$. The initial conditions for $\psi(t)$
and $\phi(t)$ are $(\psi(0), \phi(0))=(-7.6, 1)$ and $(\psi(0), \phi(0))=(-7.6,
2)$.}
\label{fig1}
\end{figure}

\begin{figure}[ht]
\centering
\includegraphics[width=5cm]{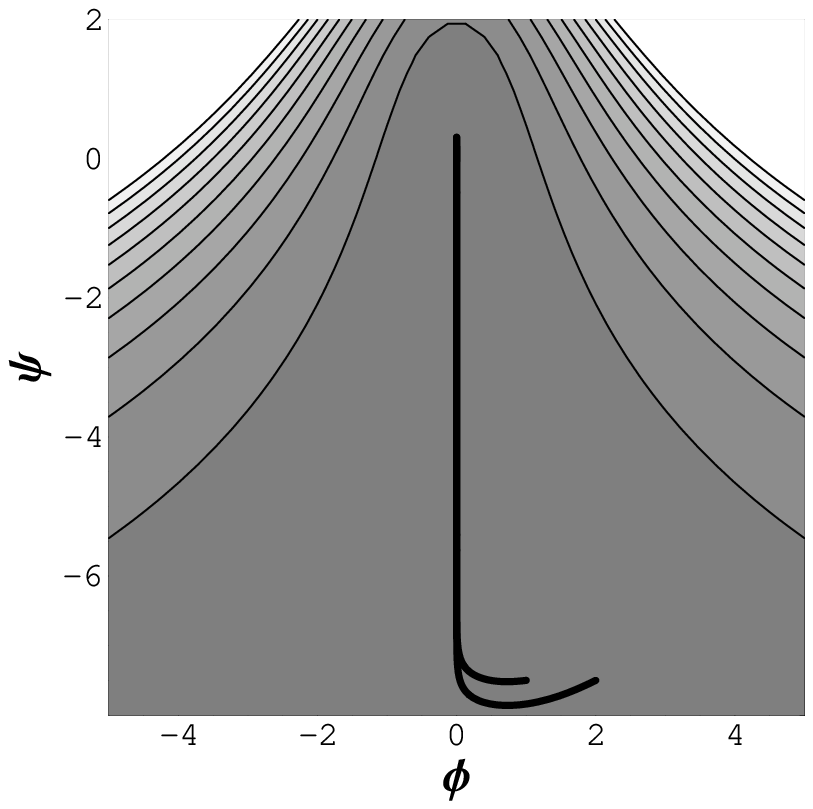}
\includegraphics[width=5cm]{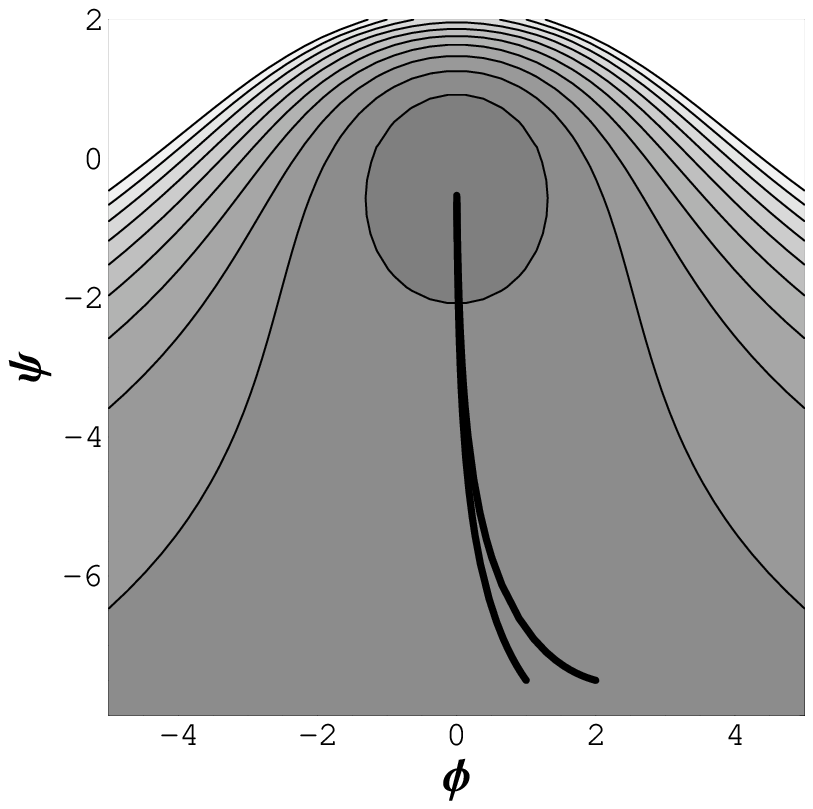}
\includegraphics[width=5cm]{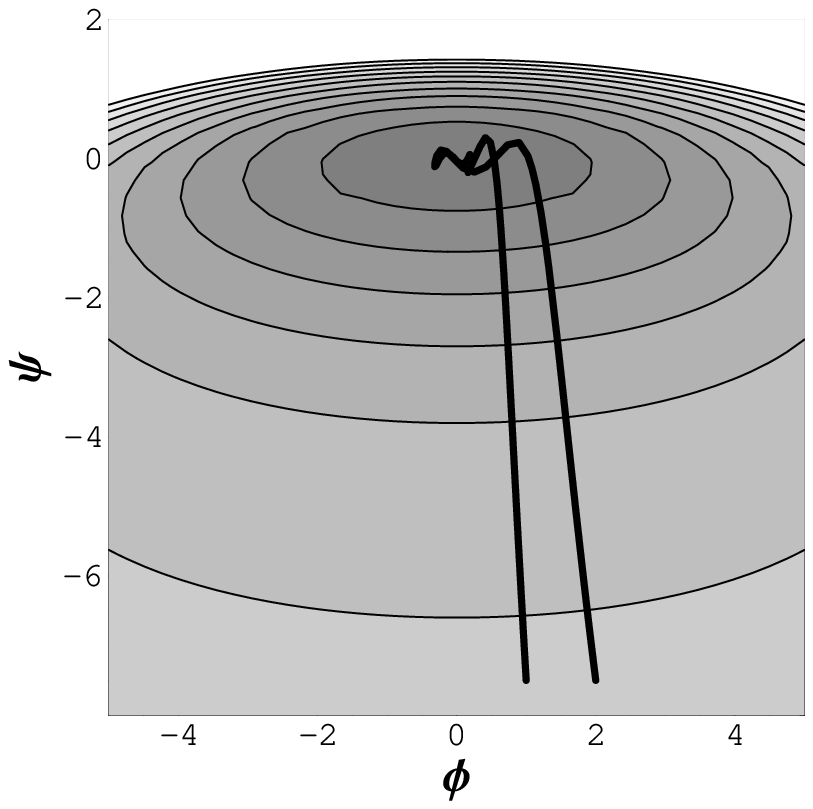}
\\
\hspace{0.5cm} (a) \hspace{4.5cm} (b) \hspace{4.5cm} (c)
\caption{Typical trajectories, on the contour plot of the potential
$U(\psi,\phi)$, for: (a) $\rho=10$, (b) $\rho=1$, (c)
$\rho=0.1$, in the case with $\gamma=1$. The initial conditions for $\psi(t)$
and $\phi(t)$ are $(\psi(0), \phi(0))=(-7.6, 1)$ and $(\psi(0), \phi(0))=(-7.6,
2)$.}
\label{fig2}
\end{figure}

\begin{figure}[ht]
\centering
\includegraphics[width=5cm]{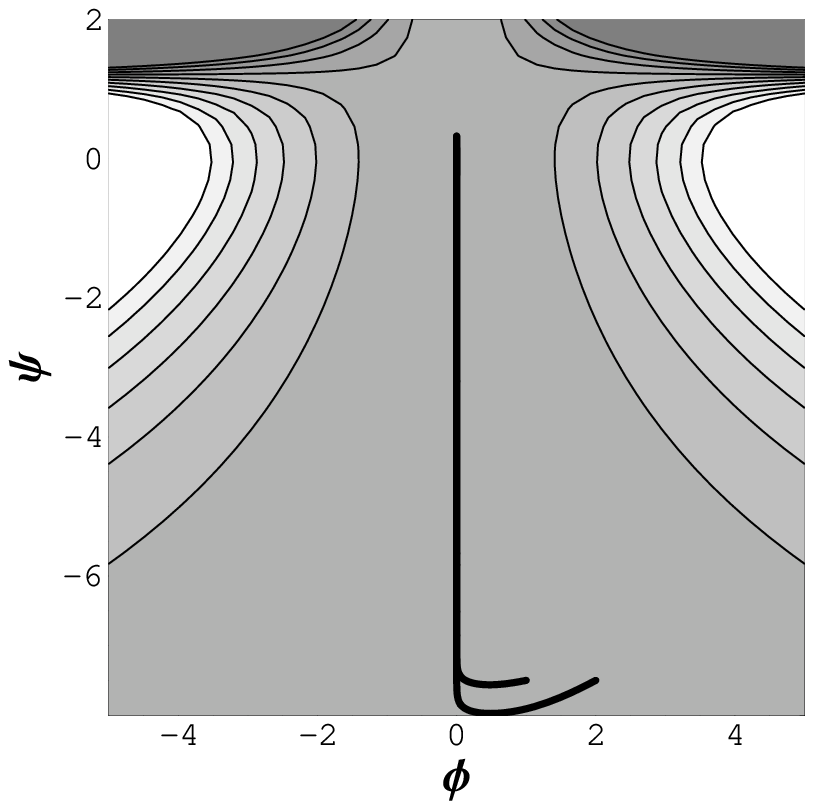}
\includegraphics[width=5cm]{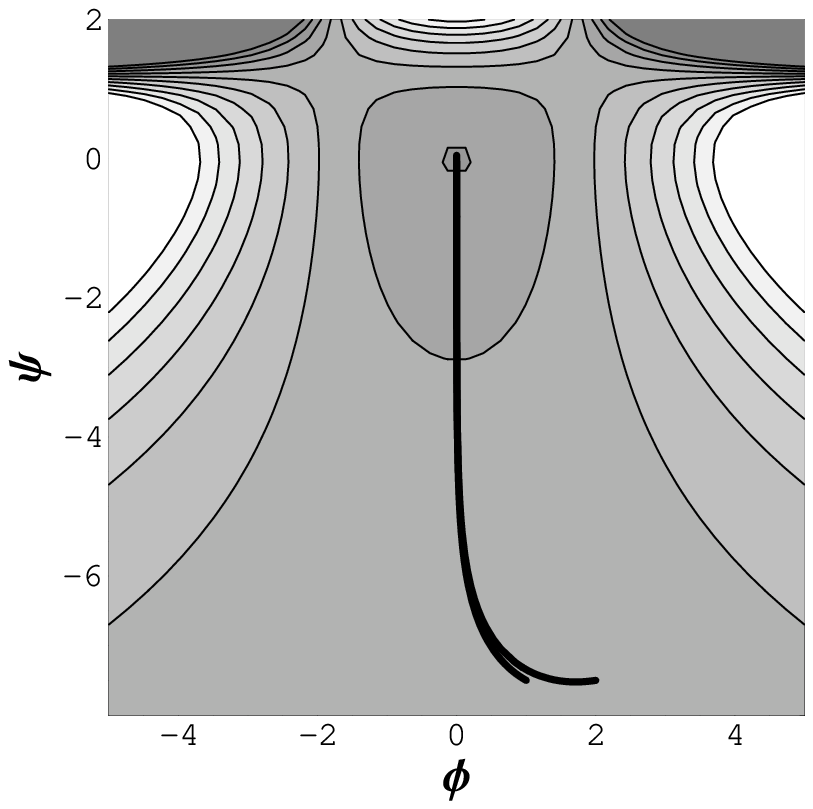}
\includegraphics[width=5cm]{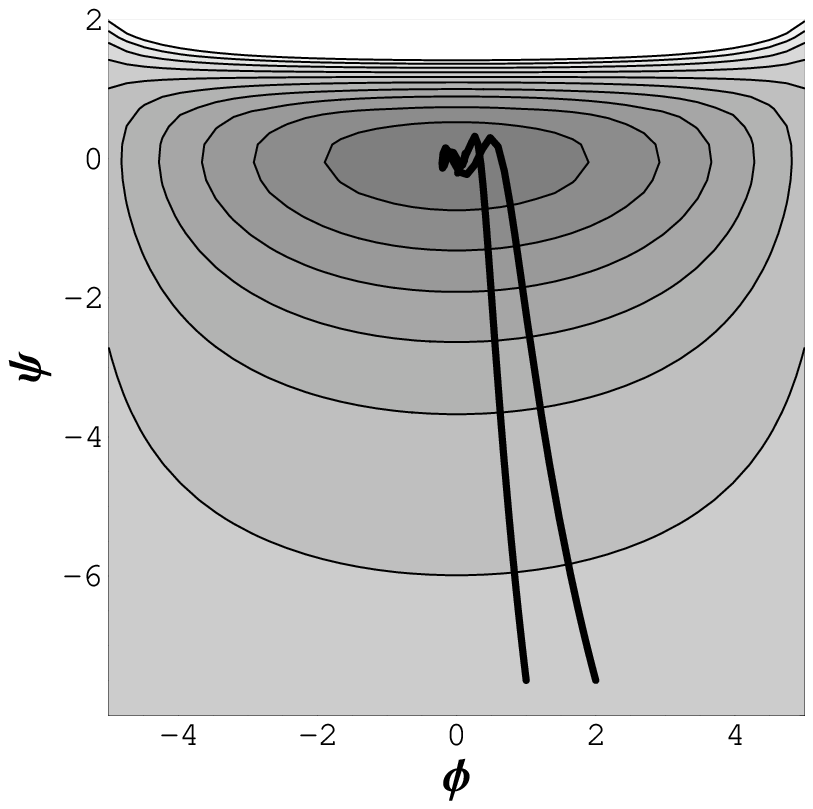}
\\
\hspace{0.5cm} (a) \hspace{4.5cm} (b) \hspace{4.5cm} (c)
\caption{Typical trajectories, on the contour plot of the potential
$U(\psi,\phi)$, for: (a) $\rho=10$, (b) $\rho=1$, (c)
$\rho=0.1$, in the case with $\gamma=2$. The initial conditions for $\psi(t)$
and $\phi(t)$ are $(\psi(0), \phi(0))=(-7.6, 1)$ and $(\psi(0), \phi(0))=(-7.6,
2)$.}
\label{fig3}
\end{figure}

In FIG.~\ref{fig1}, we show typical trajectories of the scalar fields
for $\rho=10, 1, 0.1$, in the case with the parameter $\gamma$ equals to zero.
Similar plots are shown in FIG.~\ref{fig2} and FIG.~\ref{fig3}, in the cases with
$\gamma=1$ and $\gamma=2$, respectively.
In each plots, the initial conditions for $\psi(t)$ and $\phi(t)$ are
$(\psi(0), \phi(0))=(-7.6, 1)$ and $(\psi(0), \phi(0))=(-7.6, 2)$,
and $\dot{\psi}(0)=\dot{\phi}(0)=0$ in all the plots, where the dot denotes the
time derivative.

For $\rho=10$, the value of $\phi$ first approaches zero and the scalaron
$\psi$ evolves to the potential minimum. This behavior realizes 
the Starobinsky inflation, because $V(0)=0$ reduces the potential $U$ to the
potential of the original $R^2$ model. For
$\rho=0.1$, the rapid evolution of
$\psi$ to the minimum is remarkable. For $\rho=1$, the evolution of two fields
shows intermediate behavior in the case with $\gamma=0$. For larger values of
$\gamma$, the value of $\phi$ evolves to zero more rapidly.
Thus, we conclude that there is a larger region of the initial conditions
for the Starobinsky-type inflation in the case with a larger $\gamma$.

Note that the possibility of inflation from the slow-roll of $\phi$ for a small
$\rho$ will be discussed in the last part of Sec.~\ref{qc}.

%%%%%%%%%%%%%%%%%%%%%%%%%%%
% 1
%%%%%%%%%%%%%%%%%%%%%%%%%%%
%\begin{wrapfigure}{r}{5cm}
%\end{wrapfigure}
%%%%%%%%%%%%%%%%%%%%%%%%%%%

%%%%%%%%%%%%%%%%%%%%%%%%%%%
% 2
%%%%%%%%%%%%%%%%%%%%%%%%%%%
%\begin{wrapfigure}{r}{5cm}
%\end{wrapfigure}
%%%%%%%%%%%%%%%%%%%%%%%%%%%

%%%%%%%%%%%%%%%%%%%%%%%%%%%
% 3
%%%%%%%%%%%%%%%%%%%%%%%%%%%
%\begin{wrapfigure}{r}{5cm}
%\end{wrapfigure}
%%%%%%%%%%%%%%%%%%%%%%%%%%%

Before closing this section, we write down the slow-roll parameter according
to Ref.~\cite{CEOS} here. They are, in our present notation,
\begin{eqnarray}
\epsilon&=&\frac{U_\psi^2+U_\phi^2}{U^2}\,,
\\
\eta_{\sigma\sigma}&=&\frac{2(\dot{\psi}^2U_{\psi\psi}+\dot{\phi}^2U_{\phi\phi}
+2\dot{\psi}\dot{\phi}U_{\psi\phi})}{(\dot{\psi}^2+\dot{\phi}^2)U}\,,
\end{eqnarray}
where $U_\psi\equiv\frac{\partial U}{\partial \psi}$, {etc.}~and the dot denotes
the time derivative. Further, if we assume that the scalar potential $V(\phi)$ is
slow-roll type, the analysis on approximate solutions will be identical to the
result of  Cruz-Dombriz {\it et al.}~\cite{CEOS}, independently to the value of
$\gamma$ (thus, we do not repeat it further).

%%%%%%%%%%%%%%%%%%%%%%%%%%%%%%%%%%%%%%%%%%%%%%%%%%%%%%%%%%%%%%%%%%%%%%%%%%%
%%%%%%%%%%%%%%%%%%%%%%%%%%%%%%%%%%%%%%%%%%%%%%%%%%%%%%%%%%%%%%%%%%%%%%%%%%%
%%%%%%%%%%%%%%%%%%%%%%%%%%%%%%%%%%%%%%%%%%%%%%%%%%%%%%%%%%%%%%%%%%%%%%%%%%%
\section{quantum cosmology of the extended $R^2$ model}
\label{qc}
%%%%%%%%%%%%%%%%%%%%%%%%%%%%%%%%%%%%%%%%%%%%%%%%%%%%%%%%%%%%%%%%%%%%%%%%%%%
%%%%%%%%%%%%%%%%%%%%%%%%%%%%%%%%%%%%%%%%%%%%%%%%%%%%%%%%%%%%%%%%%%%%%%%%%%%
%%%%%%%%%%%%%%%%%%%%%%%%%%%%%%%%%%%%%%%%%%%%%%%%%%%%%%%%%%%%%%%%%%%%%%%%%%%
We now turn to the quantum cosmology of our $K$-essentially modified model.
Quantum cosmology of the Starobinsky model has already been discussed by many
authors \cite{Vilenkin,HL,HW,MMS2}. Here, we shall quantize our model along with
the most standard minisuperspace method of ones used by them. 

We shall consider the metric of the form
\begin{equation}
ds^2=g_{\mu\nu}dx^\mu
dx^\nu=\frac{1}{K}a^2(t)(-dt^2+d\tilde{\Omega}_3^2)\,,
\label{met}
\end{equation}
where $d\tilde{\Omega}^2_3$ is 
the line element on a
three-dimensional maximally-symmetric manifold, whose Ricci tensor is given by
\begin{equation}
\tilde{R}_j^i=2k\delta_j^i\,,
\end{equation}
where $i, j=1,2,3$ and $k$ is a constant, which has been normalized to
$0$, $\pm 1$.
$K$ is a constant which will be used to arrange the cosmological Lagrangian
into a canonical form. 
We also assume that the scalar field $\phi$ is homogeneous on the
three dimensional manifold, i.e., $\phi=\phi(t)$.

The scalar curvature of the spacetime is computed with the metric (\ref{met}) as
\begin{equation}
R=6K\left(\frac{\ddot{a}}{a^3}+\frac{k}{a^2}\right)\,,
\end{equation}
where the dot denotes the derivative with respect to $t$.
Thus, 
the effective cosmological Lagrangian $L$, which expresses the action as
 $S=\int dt\,L$, can be
obtained as%
\footnote{Here, we adopt the normalization of the spatial volume by considering
the space as $S^3$ $(k=1)$. Even for other values of $k$, we can still use this
convention because the normalization can be absorbed into the redefinition of
$\alpha$ and $\beta$.}
\begin{eqnarray}
L&=&
\frac{2\pi^2\alpha}{K}\left[6\left(-\dot{a}^2+ka^2\right)
+\frac{1}{2}a^{2}\dot{\phi}^2-\frac{a^4}{K}V(\phi)\right]
\nonumber \\& &\qquad\qquad\quad
+\frac{2\pi^2\beta}{2}\left[6\left(\frac{\ddot{a}}{a}+k\right)
+\frac{1}{2}\dot{\phi}^2-\gamma \frac{a^2}{K}V(\phi)\right]^2\nonumber \\
&=&\frac{2\pi^2\alpha}{K}\left[6\left(-\dot{a}^2+ka^2\right)
+\frac{1}{2}a^{2}\dot{\phi}^2-\frac{a^4}{K}V(\phi)\right]+
\frac{1}{36\pi^2\beta}\bar{Q}^2
(a,\dot{a},\ddot{a},\phi,\dot{\phi})\,,
\end{eqnarray}
where we added the standard Gibbons--Hawking--York boundary term \cite{York,GH},
and further we defined
\begin{equation}
\bar{Q}\equiv 6\pi^2\beta\left[6\left(\frac{\ddot{a}}{a}+k\right)
+\frac{1}{2}\dot{\phi}^2-\gamma \frac{a^2}{K}V(\phi)\right]\,.
\label{qbar}
\end{equation}

We now select a specific value for $K$, as $K=24\pi^2\alpha$, for convenience in
the present section. We also consider an equivalent Lagrangian $L'$ as follows:
\begin{eqnarray}
L'(a,\dot{a},Q,\dot{Q},\phi,\dot{\phi})&\equiv&
L-\frac{1}{36\pi^2\beta}(Q-\bar{Q})^2-2\frac{d}{dt}\left(\frac{\dot{a}}{a}Q
\right)\nonumber \\ 
&=&\frac{1}{12}\left[-6\dot{a}^2+\frac{1}{2}a^2\dot{\phi}^2\right]+
\frac{1}{3}\left[6\frac{\dot{a}^2}{a^2}Q-6
\frac{\dot{a}}{a}\dot{Q}+\frac{1}{2}
Q\dot{\phi}^2\right]-{\cal U}(a,Q,\phi)
\nonumber \\ 
&=&\left[-\frac{1}{2}\left(a+\frac{4Q}{a}\right)^{\cdot}\dot{a}
+\frac{1}{24}(
a^2+4Q)\dot{\phi}^2\right]-{\cal U}(a,Q,\phi)\,,
\end{eqnarray}
with
\begin{equation}
{\cal
U}(a,Q,\phi)\equiv-\frac{k}{2}(a^2+2Q)+\frac{1}{36\pi^2\beta}Q^2+
\frac{1}{288\pi^2\alpha}(a^4+
4\gamma a^2 Q)V(\phi)\,.
\label{epot}
\end{equation}
We have introduced a new variable $Q$, which can be integrated out trivially as a
Gaussian integral.

Here, one can use new coordinates $x$ and $y$:
\begin{equation}
x=\frac{2Q}{a}\,,\quad
y=a+\frac{2Q}{a}\,.
\end{equation}
Additionally, one can define
$\varphi\equiv\phi/(2\sqrt{3})$ to simplify the form of the Lagrangian.
Then, the following Lagrangian is obtained:
\begin{equation}
L'=\frac{1}{2}\left(\dot{x}^2-\dot{y}^2\right)+\frac{1}{2}
(y^2-x^2)\dot{\varphi}^2-{\cal U}(x,y,\varphi)\,,
\label{xy1}
\end{equation}
where
\begin{equation}
{\cal
U}(x,y,\varphi)\equiv-\frac{k}{2}(y^2-x^2)+\frac{1}{144\pi^2\beta}x^2(y-x)^2+
\frac{1}{288\pi^2\alpha}(y-x)^3[y+(2\gamma-1)x]V(2\sqrt{3}\varphi)\,.
\label{xy2}
\end{equation}
Furthermore, using a set of variables $r$ and $\theta$ defined as
\begin{equation}
x=r\sinh\theta\,,\quad y=r\cosh\theta\,,
\end{equation}
we obtain the expression
\begin{equation}
L'=\frac{1}{2}\left[-\dot{r}^2+r^2(\dot{\theta}^2+\dot{\varphi}^2)\right]-{\cal
U}(r,\theta,\varphi)\,,
\end{equation}
with
\begin{equation}
{\cal
U}(r,\theta,\varphi)=-\frac{kr^2}{2}+\frac{r^4}{576\pi^2\beta}(1-e^{-2\theta})^2+
\frac{r^4}{288\pi^2\alpha}\left[\gamma
e^{-2\theta}-(\gamma-1)e^{-4\theta}\right]V(2\sqrt{3}\varphi)\,.
\end{equation}
Comparing with the expression in the previous section,
%correspondence of $\theta\leftrightarrow\psi$ is apparent up to normalization.
%R1
we find that the correspondences are $\phi=2\sqrt{3}\varphi$ and
$\psi=-2\sqrt{3}\theta$.

One can then treat $r$, $\theta$, and $\varphi$ as independent coordinates and
define their canonical momenta in the usual way:
\begin{equation}
P_r=\frac{\partial L'}{\partial \dot{r}}=-\dot{r}\,,\quad
P_\theta=\frac{\partial L'}{\partial \dot{\theta}}=r^2\dot{\theta}\,,\quad
P_\varphi=\frac{\partial L'}{\partial \dot{\varphi}}=r^2
\dot{\varphi}\,.
\end{equation}
One can then define the classical Hamiltonian by the normal Legendre
transformation:
\begin{equation}
H=P_r\dot{r}+P_\theta\dot{\theta}+P_\varphi\dot{\varphi}-L'%\nonumber \\
=\frac{1}{2}\left[-P_r^2+\frac{1}{r^2}\left(P_\theta^2+P_\varphi^2\right)\right]+{\cal
U}(r,\theta,\varphi)\,.
\end{equation}

The quantization scheme requires replacing $P_r$, $P_\theta$, and $P_\varphi$
by $-i\frac{\partial}{\partial r}$, $-i\frac{\partial}{\partial \theta}$,
and $-i\frac{\partial}{\partial \varphi}$, respectively.
This replacement results in the derivation of the Hamiltonian operator $\hat{H}$.
The Wheeler--De Witt (WDW) equation reads $\hat{H}\Psi=0$, i.e.,
\begin{equation}
\left[\frac{1}{r^s}\frac{\partial}{\partial
r}r^s\frac{\partial}{\partial
r}-\frac{1}{r^2}\left(\frac{\partial^2}{\partial\theta^2}+
\frac{\partial^2}{\partial\varphi^2}\right)+2{\cal
U}(r,\theta,\varphi)\right]\Psi(r,\theta,\varphi)=0\,,
\label{WDW}
\end{equation}
where $\Psi$ is the wave function of the universe. Here, we took the 
factor ordering ambiguity into consideration through the factor $r^s$.
%Eq.~(\ref{WDW}) is a three dimensional wave equation with potential
%$2{\cal U}$. 

To solve the WDW equation, we have to specify the
boundary conditions.
In this section, we restrict ourselves on the case of the closed universe, $k=1$.
In this case, the curvature-dependent term in ${\cal U}$ makes a finite potential
barrier between $r\sim 0$ and $r\gg \sqrt{\beta}$.
We first consider the behavior of the wave function in the vicinity of $r=0$.
In the beginning of the universe, or the initial state of the universe, $r$ should
be very small. For $r\ll \sqrt{\beta}$, we can neglect the terms of order of $r^4$,
i.e.,
\begin{equation}
\left[\frac{1}{r^s}\frac{\partial}{\partial
r}r^s\frac{\partial}{\partial
r}-\frac{1}{r^2}\left(\frac{\partial^2}{\partial\theta^2}+
\frac{\partial^2}{\partial\varphi^2}\right)-r^2\right]\Psi(r,\theta,\varphi)=0\,.
\label{smallq}
\end{equation}
For small $r$, both authors of \cite{Vilenkin,MMS2} took $s=0$ and assumed that
$\Psi$ is independent of the other variables. If we also assume that the wave
function is almost independent of
$\theta$ and $\varphi$, the solution of (\ref{smallq}) is the same as the solution
$\Psi\propto\sqrt{r}K_{1/4}(r^2/2)$, where $K_\nu(z)$ is the modified Bessel
function of the second kind \cite{Vilenkin,MMS2}. We here point out that if $s=1$,
the solution of  (\ref{smallq}) is given by 
\begin{equation}
\Psi\propto \int\int A(l,m)\,K_{i\sqrt{l^2+m^2}/2}(r^2/2) e^{il\theta+im\varphi} dl
dm\,,
\end{equation}
where $A(l,m)$ represents for the amplitude of each elementary wave.
In both cases, we find 
\begin{equation}
\Psi\propto e^{-r^2/2} \mbox{~~for large~} r\,.
\label{K}
\end{equation}
The problem of the ordering is also discussed in Appendix \ref{AA}.

For the region of $0<r<\sqrt{\beta}$, we use the WKB approximation, i.e.,
we want to find the solution
that has the approximate form $\Psi=A e^{-B}$.
The lowest order equation tells%
\footnote{In comparison with \cite{Vilenkin,MMS2}, absence of small parameter
here is an artifact; the replacement $r\rightarrow r/\sqrt{\lambda}$ bring about
the parameter $\lambda$. Thus, we conserve the present expressions here.}
\begin{equation}
\left(\frac{\partial B}{\partial
r}\right)^2-\frac{1}{r^2}\left[\left(\frac{\partial B}{\partial\theta}\right)^2+
\left(\frac{\partial B}{\partial\varphi}\right)^2\right]+2{\cal
U}(r,\theta,\varphi)=0\,.
\label{WKB}
\end{equation}
Note that the factor ordering does not affect the expression because we consider
$B$ as a smooth function at relatively large $r$. If the terms proportional to
$\left(\frac{\partial B}{\partial\theta}\right)^2$ and
$\left(\frac{\partial B}{\partial\varphi}\right)^2$ can be neglected
\cite{Vilenkin,MMS2}, we find
\begin{equation}
\frac{\partial B}{\partial
r}=\pm\sqrt{-2{\cal
U}(r,\theta,\varphi)}=\pm r\left[1-r^2{\cal V}(\theta,\varphi)\right]^{1/2}\,,
\end{equation}
where
\begin{equation}
{\cal
V}(\theta,\varphi)\equiv\frac{1}{144\pi^2\alpha}\left[\frac{1}{2\beta'}(1-e^{-2\theta})^2
+\left[\gamma
e^{-2\theta}-(\gamma-1)e^{-4\theta}\right]V(2\sqrt{3}\varphi)\right]\,.
\end{equation}
Therefore, we obtain $B$ in the WKB approximation as
\begin{equation}
B_{\pm}=\frac{\pm 1}{3{\cal V}(\theta,\varphi)}\left\{1-[1-r^2{\cal
V}(\theta,\varphi)]^{3/2}\right\}\,.
\end{equation}
One can see that $B_+\rightarrow r^2/2$ for $r\rightarrow 0$.
This agrees the asymptotic behavior of the wave function for a small $r$,
(\ref{K}), if we consider the tunneling wave function \`a la
Vilenkin \cite{Vilenkin}.

The wave function after tunneling, $r\gg\sqrt{\beta}$, can be obtained by
analytic continuation, thus we get
\begin{equation}
\Psi_{\pm}\propto\exp\left[\mp\frac{1}{3{\cal
V}(\theta,\varphi)}\left\{1+i[1-r^2{\cal
V}(\theta,\varphi)]^{3/2}\right\}\pm\frac{i\pi}{4}\right]\,.
\end{equation}

The `tunneling' wave function \`a la Vilenkin is proportional to $\Psi_+$, while
the wave function with the `no-boundary' boundary condition is proportional
to
$e^{\frac{2}{3{\cal V}}}\Psi_++\Psi_-$ \cite{MMS2}.
Thus, the distribution is given by \cite{MMS2}
\begin{eqnarray}
& &|\Psi|^2\propto\exp\left[-\frac{2}{3{\cal V}(\theta,\varphi)}\right]\,
\quad\mbox{for the `tunneling' boundary condition}\,,\\
& &|\Psi|^2\propto\exp\left[+\frac{2}{3{\cal V}(\theta,\varphi)}\right]\,
\quad\mbox{for the `no-boundary' boundary condition}\,.
\end{eqnarray}

The no-boundary wave function for a two-field inflation model was studied by Hwang
{\it et al.}~\cite{HKY}. 
Their study can be applied to our model with the canonical scalar
kinetic terms, provided that the potential ${\cal V}(\theta,\varphi)$
is approximated by $\frac{1}{2}{\cal V}_{\theta\theta}\theta^2+\frac{1}{2}{\cal
V}_{\varphi\varphi}\varphi^2$ (i.e., it is assumed that $V(0)=V'(0)=0$). According
to Ref.~\cite{HKY}, if
${\cal
V}_{\varphi\varphi}\ll{\cal
V}_{\theta\theta}$, cosmic inflation can occur with the number of $e$-foldings
\begin{equation}
{\cal N}\approx \frac{{\cal V}_{\theta\theta}}{{\cal
V}_{\varphi\varphi}}=\frac{\alpha}{\beta V''(0)}\,,
\end{equation}
by analysis using an approximation ${\cal V}\approx \frac{1}{2}{\cal
V}_{\theta\theta}\theta^2+{\cal V}_0$, where ${\cal V}_0=\frac{1}{2}{\cal
V}_{\varphi\varphi}\varphi^2\approx const$~\cite{HKY}. Therefore, as an
inflationary two-field model, our model can work well.

%R1
Here, we confirmed that our model admits no-boundary wave function which
can provide an appropriate initial condition for the two-field inflation model
in the preceding study of Hwang {\it
et al.}~\cite{HKY}. Although the quantum effect on evolution of
the universe is also an important subject to study, we think that it is beyond our
present scope and should be left aside for future work.
%R1

Before closing this section, we place a comment on another method of analysis on
the wave function of
$r\ll\sqrt{\beta}$ (which is not limited on the extended model).
If one removes the assumption that $\Psi$ is independent of $\theta$,
we can write the WDW equation in terms of $x$ and $y$
from Eqs.~(\ref{xy1}) and (\ref{xy2}):
\begin{equation}
\left[-\frac{\partial^2}{\partial x^2}+x^2+\frac{\partial^2}{\partial
y^2}-y^2\right]\Psi(x,y)=0\,,
\end{equation}
and the wave packet solution is \cite{Kiefer}
\begin{equation}
\Psi(x,y)=\sum_n
A_n\frac{H_n(x)H_n(y)}{2^nn!}\exp\left[-\frac{1}{2}(x^2+y^2)\right]\,,
\end{equation}
where $A_n$ is amplitudes.
The asymptotic behavior at $r\rightarrow\infty$ is
$\Psi\sim \exp[-\frac{1}{2}r^2]$ can be obtained for $\theta\sim 0$ (i.e., $x\sim
0$).

%%%%%%%%%%%%%%%%%%%%%%%%%%%%%%%%%%%%%%%%%%%%%%%%%%%%%%%%%%%%%%%%%%%%%%%%%%%
%%%%%%%%%%%%%%%%%%%%%%%%%%%%%%%%%%%%%%%%%%%%%%%%%%%%%%%%%%%%%%%%%%%%%%%%%%%
%%%%%%%%%%%%%%%%%%%%%%%%%%%%%%%%%%%%%%%%%%%%%%%%%%%%%%%%%%%%%%%%%%%%%%%%%%%
\section{modified positive-definite action}
\label{pd}
%%%%%%%%%%%%%%%%%%%%%%%%%%%%%%%%%%%%%%%%%%%%%%%%%%%%%%%%%%%%%%%%%%%%%%%%%%%
%%%%%%%%%%%%%%%%%%%%%%%%%%%%%%%%%%%%%%%%%%%%%%%%%%%%%%%%%%%%%%%%%%%%%%%%%%%
%%%%%%%%%%%%%%%%%%%%%%%%%%%%%%%%%%%%%%%%%%%%%%%%%%%%%%%%%%%%%%%%%%%%%%%%%%%

Positive-definite action for gravity was conjectured by Horowitz \cite{Horowitz}
about three decades ago.
Under some appropriate conditions, this theory can be considered as the
high-curvature limit of $R+\beta R^2/2$ theory.
An extension of the model with a scalar field is defined by the following action:
\begin{equation}
S_{S}=\frac{\beta}{2}\int
d^4x\,\sqrt{-g}\,\left[R-\frac{1}{2}(\nabla\phi)^2-
\gamma V(\phi)\right]^2\,.
\end{equation}
By a similar method to that seen in Sec.~\ref{cc}, the equivalent action at
classical level is obtained as
\begin{equation}
S=\int d^4x\sqrt{-g}\,\left\{\beta\chi\left[R-\frac{1}{2}(\nabla\phi)^2-
\gamma V(\phi)\right]-\frac{\beta}{2}\chi^2\right\}\,.
\end{equation}
An appropriate Weyl transformation with the apparent auxiliary field $\chi$ is
attained by the transformation
% $\tilde{g}_{\mu\nu}$ which satisfies
%$\sqrt{-g}\beta\chi R=\sqrt{-\tilde{g}}\beta\tilde{R}+\cdots$,
%where $\tilde{R}$ is the Ricci scalar constructed from $\tilde{g}_{\mu\nu}$.
$g_{\mu\nu}=\chi^{-1}\tilde{g}_{\mu\nu}$.
Then, we obtain
\begin{equation}
S=\int d^4x\sqrt{-\tilde{g}}\,\,\beta\left[
\tilde{R}-\frac{1}{2}(\tilde{\nabla}\phi)^2-
\frac{1}{2}(\tilde{\nabla}\psi)^2-
e^{\frac{1}{\sqrt{3}}\psi}\gamma V(\phi)-\frac{1}{2}\right]\,,
\label{R2}
\end{equation}
where we defined $\psi\equiv-\sqrt{3}\ln\chi$.

There appears the cosmological constant as the last term in the Lagrangian in
(\ref{R2}). In our present model, however, the scalar potential which can be
arbitrarily chosen is included in the Lagrangian. Thus, a general two-field model
for cosmic acceleration can be constructed in our scheme.

Classical cosmological solutions are known to be obtained analytically in some
cases, but we leave the derivation of solutions for the next section, where we
exhibit solvable pure $R^p$ models. 

%%%%%%%%%%%%%%%%%%%%%%%%%%%%%%%%%%%%%%%%%%%%%%%%%%%%%%%%%%%%%%%%%%%%%%%%%%%
%%%%%%%%%%%%%%%%%%%%%%%%%%%%%%%%%%%%%%%%%%%%%%%%%%%%%%%%%%%%%%%%%%%%%%%%%%%
%%%%%%%%%%%%%%%%%%%%%%%%%%%%%%%%%%%%%%%%%%%%%%%%%%%%%%%%%%%%%%%%%%%%%%%%%%%

Quantum cosmology of $R^2$ gravity has been investigated in many papers, including
\cite{Horowitz,Schmidt1,Schmidt2}. We study our extended model by the standard
method used by them. As in the previous section, we obtain the effective
cosmological Lagrangian of the model:
\begin{eqnarray}
&
&L\equiv\frac{2\pi^2\beta}{2}\left[6\left(\frac{\ddot{a}}{a}+1\right)
+\frac{1}{2}\dot{\phi}^2-\gamma \frac{a^2}{K}V(\phi)\right]^2
=\frac{1}{36\pi^2\beta}\bar{Q}^2
(a,\dot{a},\ddot{a},\phi,\dot{\phi})\,,
\end{eqnarray}
where $\bar{Q}$ is defined as in Eq.~(\ref{qbar}), and other metric definitions are
the same as well in the previous section.  The equivalent Lagrangian now has the
form
\begin{eqnarray}
L'(a,\dot{a},Q,\dot{Q},\phi,\dot{\phi})&\equiv&
L-\frac{1}{36\pi^2\beta}(Q-\bar{Q})^2-2\frac{d}{dt}\left(\frac{\dot{a}}{a}Q
\right)\nonumber \\ 
%& &=\frac{1}{3}\left[6\frac{\dot{a}^2}{a^2}Q-6 
%\frac{\dot{a}}{a}\dot{Q}+\frac{1}{2} 
%Q\dot{\phi}^2\right]-{\cal U}(a,Q,\phi)\nonumber\\ 
&=&
-2\left(\frac{Q}{a}\right)^{\cdot}\dot{a}
+\frac{1}{6}Q\dot{\phi}^2-{\cal U}(a,Q,\phi)\,,
\end{eqnarray}
where
\begin{equation}
{\cal
U}(a,Q,\phi)\equiv\frac{1}{36\pi^2\beta}Q^2-2kQ+\frac{a^2}{3K}
Q\gamma V(\phi)\,.
\end{equation}
Here, one can use new coordinates $x$ and $y$:
\begin{equation}
x=\frac{Q}{a}-a\,,\quad
y=a+\frac{Q}{a}\,.
\end{equation}
We define
$\varphi\equiv\phi/(2\sqrt{3})$ to simplify the form of the Lagrangian.
Then, one can obtain
\begin{equation}
L'=\frac{1}{2}\left(\dot{x}^2-\dot{y}^2\right)+\frac{1}{2}
(y^2-x^2)\dot{\varphi}^2-{\cal U}(x,y,\phi)\,,
\end{equation}
where
\begin{equation}
{\cal
U}(x,y,\phi)\equiv-\frac{k}{2}(y^2-x^2)+\frac{1}{576\pi^2\beta}(y^2-x^2)^2+
\frac{1}{48K}(y-x)^2(y^2-x^2)\gamma V(2\sqrt{3}\varphi)\,.
\end{equation}
Using a set of variables $r$ and $\theta$ defined as
$x=r\sinh\theta$ and $y=r\cosh\theta$,
we obtain the expression
\begin{equation}
L'=\frac{1}{2}\left[-\dot{r}^2+r^2(\dot{\theta}^2+\dot{\varphi}^2)\right]-{\cal
U}(r,\theta,\varphi)\,,
\end{equation}
with
\begin{equation}
{\cal
U}(r,\theta,\varphi)=-\frac{k}{2}r^2+\frac{r^4}{576\pi^2\beta}+
\frac{1}{48K}r^4e^{-2\theta}\gamma V(2\sqrt{3}\varphi)\,.
\end{equation}
Then, we obtain the Hamiltonian
\begin{eqnarray}
H&=&\frac{1}{2}\left[-P_r^2+\frac{1}{r^2}\left(P_\theta^2+P_\varphi^2\right)\right]+{\cal
U}(r,\theta,\varphi)\,.
\end{eqnarray}
and the WDW equation
\begin{equation}
\left[\frac{1}{r}\frac{\partial}{\partial
r}r\frac{\partial}{\partial
r}-\frac{1}{r^2}\left(\frac{\partial^2}{\partial\theta^2}+
\frac{\partial^2}{\partial\varphi^2}\right)+2{\cal
U}(r,\theta,\varphi)\right]\Psi(r,\theta,\varphi)=0\,,
\end{equation}
as in the previous section (and we fixed here the ordering by $s=1$).

%%%%%%%%%%%%%%%%%%%%%%%%%%%%%%%%%%%%%%%%%%%%%%%%%%%%%%%%%%%%%%%%%%%%%%%%%%%

Let us consider the solution of the WDW equation.
The case of a closed universe is similarly analyzable as the model in the previous
section. Therefore, we here consider the case of a flat universe, $k=0$.

Moreover, if $V(\phi)\equiv 0$ or
$V(\phi)$ is negligible, the equation becomes
\begin{equation}
\left[\frac{1}{r}\frac{\partial}{\partial
r}r\frac{\partial}{\partial
r}-\frac{1}{r^2}\left(\frac{\partial^2}{\partial\theta^2}+
\frac{\partial^2}{\partial\varphi^2}\right)+\frac{r^4}{288\pi^2\beta}
\right]\Psi(r,\theta,\varphi)=0\,.
\end{equation}
The solution of this equation can be expressed by the form of superposition:
\begin{eqnarray}
\Psi&=&\int\int \left[
A(l,m) J_{i\frac{\sqrt{l^2+m^2}}{3}}(r^3/(36\sqrt{2\beta}\pi))\right.
\nonumber \\
& &\qquad\quad\left.+
B(l,m) J_{-i\frac{\sqrt{l^2+m^2}}{3}}(r^3/(36\sqrt{2\beta}\pi))\right]
e^{il\theta+im\varphi} dl dm\,,
\label{ps}
\end{eqnarray}
where $J_\nu(z)$ is the Bessel function and $A(l,m)$ and $B(l,m)$ are amplitudes
for each elementary wave. Because of the absence of potential wall due to the
curvature and the scalar potential, the behavior of the wave function is generally
oscillatory in the direction of
$r$.
Since there is no tunneling, we assume an appropriate wave packet form
in the beginning of the universe \cite{Kiefer,KNw,Kiefer1} to study
further.

As seen here, it is known that some limited cases can be solved exactly.
In the next section, we pursue the solvable model of $K$-essentially modified
pure $R^p$ gravity, where $p$ is a rational number.

%%%%%%%%%%%%%%%%%%%%%%%%%%%%%%%%%%%%%%%%%%%%%%%%%%%%%%%%%%%%%%%%%%%%%%%%%%%
%%%%%%%%%%%%%%%%%%%%%%%%%%%%%%%%%%%%%%%%%%%%%%%%%%%%%%%%%%%%%%%%%%%%%%%%%%%
%%%%%%%%%%%%%%%%%%%%%%%%%%%%%%%%%%%%%%%%%%%%%%%%%%%%%%%%%%%%%%%%%%%%%%%%%%%
\section{soluble models for $K$-essentially modified pure $R^p$ gravity}
\label{monomial}
%%%%%%%%%%%%%%%%%%%%%%%%%%%%%%%%%%%%%%%%%%%%%%%%%%%%%%%%%%%%%%%%%%%%%%%%%%%
%%%%%%%%%%%%%%%%%%%%%%%%%%%%%%%%%%%%%%%%%%%%%%%%%%%%%%%%%%%%%%%%%%%%%%%%%%%
%%%%%%%%%%%%%%%%%%%%%%%%%%%%%%%%%%%%%%%%%%%%%%%%%%%%%%%%%%%%%%%%%%%%%%%%%%%

In this section, we consider an extension of pure $R^p$ gravity in $D$ dimensional
spacetime. We start with the action
\begin{equation}
S_{}=\frac{\beta}{p(p-1)}\int
d^Dx\,\sqrt{-g}\,\left[R-\frac{1}{2}(\nabla\phi)^2-
\gamma V(\phi)\right]^p
\,.
\end{equation}
As in the previous sections, we can use an auxiliary field $\chi$ to obtain
classically equivalent action:
\begin{equation}
S=\beta\int
d^Dx\sqrt{-g}\,\left\{\frac{\chi^{p-1}}{p-1}\left[R-\frac{1}{2}(\nabla\phi)^2-
\gamma V(\phi)\right]-\frac{\chi^p}{p}\right\}\,.
\label{rp}
\end{equation}
We eliminate the $\chi$-dependence in front of $R$
in the action (\ref{rp}) by a Weyl transformation.
We consider a metric $\tilde{g}_{\mu\nu}$
which satisfies
$\sqrt{-g}\chi^{p-1} R=\sqrt{-\tilde{g}}\tilde{R}+\cdots$,
where $\tilde{R}$ is the Ricci scalar constructed from $\tilde{g}_{\mu\nu}$.
Now, we select
$g_{\mu\nu}=\chi^{-\frac{2(p-1)}{D-2}}\tilde{g}_{\mu\nu}$ in this time.
Then, we obtain
\begin{eqnarray}
S&=&\frac{\beta}{p-1}\int d^Dx\sqrt{-\tilde{g}}\left[
\tilde{R}-\frac{1}{2}(\tilde{\nabla}\phi)^2-
\frac{1}{2}(\tilde{\nabla}\psi)^2\right.\nonumber \\
& &\qquad\qquad\qquad\qquad\left.-
e^{\sqrt{\frac{2}{(D-1)(D-2)}}\psi}\gamma
V(\phi)-\frac{p-1}{p}e^{\sqrt{\frac{2}{(D-1)(D-2)}}\frac{p-D/2}{p-1}\psi}\right]\,.
\end{eqnarray}
Here, we defined 
$\psi\equiv-\sqrt{\frac{2(D-1)}{D-2}}(p-1)\ln\chi$.
Note that the scalar field $\phi$ has a canonical kinetic term again in the model. 

We expect that the exact solutions of simple models are useful to reveal
some subtle features of the cosmological dynamical system
\cite{MP,PM,TW,Ohta2,LMPX,Kaloper,CGG,Roy,ANL,ALNW,KKST}. Here, we investigate the
cases with specific parameters, where the exact classical and quantum cosmological
solutions can be obtained.

For this purpose, we restrict ourselves on a flat $D$-dimensional spacetime and
choose the metric as
\begin{equation}
d\tilde{s}^2=-e^{2n(t)}dt^2+e^{2\tilde{a}(t)}\sum_{i=1}^{D-1}(dx^i)^2\,.
\end{equation}
From this metric, the scalar curvature can be calculated as
\begin{equation}
\tilde{R}=e^{-2n}[2(D-1)(\ddot{\tilde{a}}-\dot{n}\dot{\tilde{a}})+
D(D-1)\dot{\tilde{a}}^2]\,.
\end{equation}
Fixing a gauge $n(t)=(D-1)\tilde{a}(t)$, one can find that the cosmological
Lagrangian can be rewritten as
\begin{eqnarray}
L=& &-(D-1)(D-2)\dot{\tilde{a}}^2+\frac{1}{2}\dot{\phi}^2+
\frac{1}{2}\dot{\psi}^2\nonumber \\
& &-
e^{2(D-1)\tilde{a}+\sqrt{\frac{2}{(D-1)(D-2)}}\psi}\gamma
V(\phi)-\frac{p-1}{p}e^{2(D-1)\tilde{a}+\sqrt{\frac{2}{(D-1)(D-2)}}
\frac{p-D/2}{p-1}\psi}\,,
\label{LLL}
\end{eqnarray}
apart from an overall normalization which differs from the previous one.

We find that exact solutions can be obtained for the system governed by the
Lagrangian (\ref{LLL}) in the following two cases:
A. $p=D/2$ and $\gamma V(\phi)=0$, B. $p=1-1/(2D)$ and $\gamma
V(\phi)=\frac{1}{2}f^2
\exp[\sqrt{\frac{2D}{D-1}}g\phi]$.
The case of $p=2$ and $D=4$ seen in the previous section belongs to the case A.
We will exhibit classical and quantum exact solutions in both cases A and B.

Note that the equivalence of the higher order theory and the reduced theory
by using an auxiliary field is classically valid, while the equivalence of them in
quantum physics is not always clear.
As seen in previous sections, quantum $R^2$ gravity can be rewritten by adding
quadratic term including a new variable. This redefinition is equivalent to adding
a Gaussian integration in view of path integral. For a general power $p$, similar
addition of a new variable may cause a problem of measure in path integral.
Nevertheless, we consider the quantization of the system with an `auxiliary' field
as an `effective' theory, which can grasp some feature and tendency of 
behavior of dynamical variables in the physical system.

Now, we show the solutions in solvable cases.

%%%%%%%%%%%%%%%%%%%%%%%%%%%%%%%%%%%%%%%%%%%%%%%%%%%%%%%%%%%%%%%%%%%%%%%%%%%
%%%%%%%%%%%%%%%%%%%%%%%%%%%%%%%%%%%%%%%%%%%%%%%%%%%%%%%%%%%%%%%%%%%%%%%%%%%
\subsection{$p=D/2$ and $\gamma V(\phi)=0$}
%%%%%%%%%%%%%%%%%%%%%%%%%%%%%%%%%%%%%%%%%%%%%%%%%%%%%%%%%%%%%%%%%%%%%%%%%%%
%%%%%%%%%%%%%%%%%%%%%%%%%%%%%%%%%%%%%%%%%%%%%%%%%%%%%%%%%%%%%%%%%%%%%%%%%%%
The first case is a natural generalization of pure $R^2$ gravity in four
dimensions. In this case, $V(\phi)\equiv 0$ is assumed and $p=D/2$ is chosen.
The cosmological Lagrangian includes two massless scalar field, aside from the
cosmological term, in this case.  Then, the Lagrangian (\ref{LLL})
becomes
\begin{equation}
L=-(D-1)(D-2)\dot{\tilde{a}}^2+\frac{1}{2}\dot{\phi}^2+
\frac{1}{2}\dot{\psi}^2-\frac{D-2}{D}e^{2(D-1)\tilde{a}}\,.
\end{equation}
In order to make it simpler, we introduce normalization and other constants as
\begin{equation}
X\equiv\sqrt{2(D-1)(D-2)}\tilde{a}\,,\quad 
\lambda\equiv\sqrt{\frac{D-1}{2(D-2)}}\,,\quad
\delta\equiv\sqrt{\frac{2(D-2)}{D}}\,.
\end{equation}
We now obtain the simplified Lagrangian
\begin{equation}
L=-\frac{1}{2}\dot{X}^2-\frac{\delta^2}{2}e^{2\lambda X}
+\frac{1}{2}\dot{\phi}^2+
\frac{1}{2}\dot{\psi}^2\,,
\label{59}
\end{equation}
and the Hamiltonian derived from the Lagrangian
\begin{equation}
H=H_X+H_\phi+H_\psi\,.
\end{equation}
Here, the separated Hamiltonians are
\begin{equation}
H_X=-\frac{1}{2}p_X^2+\frac{\delta^2}{2}e^{2\lambda X}\,,\quad H_\phi=
\frac{1}{2}p_\phi^2\,,\quad
H_\psi=\frac{1}{2}p_\psi^2\,,
\end{equation}
where $p_X=-\dot{X}$, $p_\phi=\dot{\phi}$, and
$p_\psi=\dot{\psi}$.

From the Lagrangian (\ref{59}), we find that the variables are separated and $X$
obeys the Liouville equation in one dimension.
Therefore, the separated Hamiltonians $H_X$, $H_\phi$, and $H_\psi$ become
constants 
$E_X$, $E_\phi$, and $E_\psi$, respectively, if the individual solutions are
substituted. Thus, the solutions can be written down as
\begin{eqnarray}
X(t)&=&\frac{1}{2\lambda}\ln\frac{q_X^2}{\sinh^2q_X\delta\lambda(t-t_X)}\,,
\qquad E_X=-\frac{q_X^2\delta^2}{2}\,,\\
\phi(t)&=&q_\phi(t-t_\phi)\,,\quad E_\phi=\frac{q_\phi^2}{2}\,,
\\
\psi(t)&=&q_\psi(t-t_\psi)\,,\quad E_\psi=\frac{q_\psi^2}{2}\,,
\end{eqnarray}
where $q_X$, $q_\phi$, $q_\psi$, $t_X$, $t_\phi$, and $t_\psi$ are integration
constants. Remembering that we treat the general relativistic system, the
constants should satisfy the relation
\begin{equation}
E_X+E_\phi+E_\psi=0\,.
\end{equation}
Therefore, we can finally write the solution as
\begin{eqnarray}
& &e^{2(D-1)\tilde{a}(t)}=e^{2\lambda
X}=\frac{q_\phi^2+q_\psi^2}{\delta^2\sinh^2\sqrt{q_\phi^2+q_\psi^2}\lambda(t-t_X)}
\,,\nonumber \\
& &\phi(t)=q_\phi(t-t_\phi)\,,\quad
\psi(t)=q_\psi(t-t_\psi)\,.
\end{eqnarray}

One can check the solution through a special case, $q_\phi=q_\psi\rightarrow 0$.
In this limit, one finds
$e^{(D-1)\tilde{a}(t)}=\delta^{-1}\lambda^{-1}|t_X-t|^{-1}$.
If one uses the `canonical' cosmic time $\tau$, $d\tau\equiv e^n dt=
e^{(D-1)\tilde{a}(t)}dt$, one obtains $\tau=
-\delta^{-1}\lambda^{-1}\ln|t_X-t|$. Accordingly, $e^{(D-1)\tilde{a}(t)}\propto
e^{\delta\lambda\tau}$ can be found. This exponential expansion is caused by the
cosmological constant, since the scalar fields are frozen in this limit.
For general $q_\phi$ and $q_\psi$, the asymptotic behavior of the scale factor
$e^{\tilde{a}}$ can be found as: $e^{\tilde{a}}\sim e^{\delta\lambda\tau/(D-1)}$
for $\tau\rightarrow\infty$ and $e^{\tilde{a}}\sim \tau^{1/(D-1)}$
for $\tau\sim 0$.

%%%%%%%%%%%%%%%%%%%%%%%%%%%%%%%%%%%%%%%%%%%%%%%%%%%%%%%%%%%%%%%%%%%%%%%%%%%

We now turn to the study of quantum cosmology in this case.
The WDW equation becomes
\begin{equation}
\left[\frac{\partial^2}{\partial X^2}+\delta^2\,e^{2\lambda X}
-\frac{\partial^2}{\partial\phi^2}-
\frac{\partial^2}{\partial\psi^2}\right]\Psi(X,\phi,\psi)=0\,.
\end{equation}
The general solution of this wave equation is
\begin{eqnarray}
\Psi(X,\phi,\psi)&=&\int\int\left[ 
A(l,m) J_{i\sqrt{l^2+m^2}/\lambda}(\delta e^{\lambda X}/\lambda)
\right.\nonumber \\
& &\qquad\quad+\left.
B(l,m) J_{-i\sqrt{l^2+m^2}/\lambda}(\delta e^{\lambda X}/\lambda)
\right]e^{il\psi+im\phi} dl dm
\,.
\label{cs}
\end{eqnarray}
Note that $e^{\lambda X}=e^{(D-1)\tilde{a}}$.
The expression of (\ref{cs}) seems a different from the solution (\ref{ps})
in four dimensions, because of the choice of time coordinate
as well as because of the different set of variables and their normalizations.
Nevertheless, the reason of appearance of the similar type of function
can be understood easily as follows.
In the present section, we obtain $\dot{\psi}=const.$ as a consequence of the gauge
choice ($n=(D-1)\tilde{a}$). Thus, we can regard $\psi$ as a `time' in the space of
variables in this gauge. Taking the `time' axis simply so as to cross the `origin'
where $\psi\approx\theta\approx 0$, the argument $e^{3\tilde{a}}$ appears in the
Bessel function in (\ref{cs}), since $e^{3\tilde{a}}=a^3=(r/2)^3e^{-3\theta}\approx
(r/2)^3$.

%%%%%%%%%%%%%%%%%%%%%%%%%%%%%%%%%%%%%%%%%%%%%%%%%%%%%%%%%%%%%%%%%%%%%%%%%%%
%%%%%%%%%%%%%%%%%%%%%%%%%%%%%%%%%%%%%%%%%%%%%%%%%%%%%%%%%%%%%%%%%%%%%%%%%%%
\subsection{$p=1-\frac{1}{2D}$ and $\gamma
V(\phi)=\frac{1}{2}f^2\exp(\sqrt{\frac{2D}{D-1}}g\phi)$}
%%%%%%%%%%%%%%%%%%%%%%%%%%%%%%%%%%%%%%%%%%%%%%%%%%%%%%%%%%%%%%%%%%%%%%%%%%%
%%%%%%%%%%%%%%%%%%%%%%%%%%%%%%%%%%%%%%%%%%%%%%%%%%%%%%%%%%%%%%%%%%%%%%%%%%%
The second case enjoys dynamics of two scalar modes in general.
When we set $p=1-\frac{1}{2D}$ and assume $\gamma
V(\phi)=\frac{1}{2}f^2\exp(\sqrt{\frac{2D}{D-1}}g\phi)$, we obtain the following
Lagrangian 
\begin{eqnarray}
L=& &-(D-1)(D-2)\dot{\tilde{a}}^2+\frac{1}{2}\dot{\phi}^2+
\frac{1}{2}\dot{\psi}^2\nonumber \\
& &-\frac{1}{2}f^2
e^{\sqrt{\frac{2D}{D-1}}g\phi+2(D-1)\tilde{a}+\sqrt{\frac{2}{(D-1)(D-2)}}\psi}
+\frac{1}{2D-1}e^{2(D-1)\tilde{a}+(D-1)\sqrt{\frac{2(D-1)}{D-2}}
\psi}\,.
\label{ol}
\end{eqnarray}
We consider a new set of variables:
\begin{eqnarray}
x&\equiv&
\frac{1}{\sqrt{1-g^2}}
\left[\sqrt{\frac{D-1}{2D}}\left(2(D-1)\tilde{a}+\sqrt{\frac{2}{(D-1)(D-2)}}\psi
\right)+g\phi\right]\,,
\\
y&\equiv&
\frac{1}{\sqrt{1-g^2}}
\left[\phi+g\sqrt{\frac{D-1}{2D}}\left(2(D-1)\tilde{a}+\sqrt{\frac{2}{(D-1)(D-2)}}\psi
\right)\right]\,,
\\
z&\equiv&
\sqrt{\frac{D-1}{2D}}\left(2\tilde{a}+(D-1)\sqrt{\frac{2}{(D-1)(D-2)}}\psi
\right)\,,
\end{eqnarray}
and define constants:
\begin{equation}
\lambda_1\equiv\sqrt{\frac{D}{2(D-1)}}\sqrt{1-g^2}\,,
\quad
\lambda_3\equiv\sqrt{\frac{D(D-1)}{2}}\,.
\end{equation}
Then, we can obtain the simple form of the Lagrangian (\ref{ol}) as
\begin{equation}
L(x,y,z;\dot{x},\dot{y},\dot{z})=-\frac{1}{2}\dot{x}^2+\frac{1}{2}\dot{y}^2+
\frac{1}{2}\dot{z}^2-\frac{1}{2}f^2
e^{2\lambda_1x}
+\frac{1}{2D-1}e^{2\lambda_3z}\,.
\end{equation}
Then, the Hamiltonian becomes
\begin{equation}
H=H_1+H_2+H_3\,,
\label{hami}
\end{equation}
where
\begin{equation}
H_1=-\frac{1}{2}p_x^2+\frac{f^2}{2}e^{2\lambda_1 x}\,,
\quad H_2=\frac{1}{2}p_y^2\,,\quad H_3
=\frac{1}{2}p_z^2-\frac{1}{2D-1}e^{2\lambda_2 z},
\end{equation}
with $p_x=-\dot{x}$, $p_y=\dot{y}$, and $p_z=\dot{z}$. 

Because of separation of variables,
each variable can be solved by a solution of the one-dimensional Liouville
equation. The values of $H_i$ $(i=1,2,3)$ become constants $E_i$, if the
solutions are substituted into them. The exact solutions are
\begin{eqnarray}
x(t)&=&\frac{1}{2\lambda_1}\ln\frac{q_1^2}{\sinh^2q_1f\lambda_1(t-t_1)}\,,
\qquad E_1=-\frac{q_1^2f^2}{2}\qquad (f>0)\\
x(t)&=&\frac{1}{2\lambda_1}\ln\frac{q_1^2}{\sin^2q_1f\lambda_1(t-t_1)}\,,
\qquad E_1=\frac{q_1^2f^2}{2}\qquad (f>0)\\
x(t)&=&q_1(t-t_1)\,,\quad E_1=-\frac{q_1^2}{2}
\qquad (f=0)
\end{eqnarray}
\begin{equation}
y(t)=q_2(t-t_2)\,,\quad E_2=\frac{q_2^2}{2}
\end{equation}
\begin{eqnarray}
z(t)&=&\frac{1}{2\lambda_3}\ln\frac{q_3^2}{\sinh^2q_3\sqrt{(D-1/2)^{-1}}
\lambda_3(t-t_3)}\,,
\qquad E_3=\frac{q_3^2}{2D-1}\\
z(t)&=&\frac{1}{2\lambda_3}\ln\frac{q_3^2}{\sin^2q_3\sqrt{(D-1/2)^{-1}}
\lambda_3(t-t_3)}\,,
\qquad E_3=-\frac{q_3^2}{2D-1}
\end{eqnarray}
where constants $E_1$, $E_2$, and $E_3$ should satisfy
\begin{equation}
E_1+E_2+E_3=0\,.
\end{equation}
Because $E_2\ge 0$, possible combinations are ($E_1\le 0, E_3\ge 0$), ($E_1\ge 0,
E_3\le 0$), and ($E_1\le 0, E_3\le 0$).

Power-law inflation \cite{LM} can be obtain in the case ($E_1\le 0, E_3\ge 0$).
We find, in this case,
\begin{equation}
e^{(D-1)\tilde{a}}\propto
\exp\left(\frac{(D-1)^{3/2}g}{(D-2)\sqrt{2D(1-g^2)}}q t\right)
\frac{[\sinh q\sinh\theta 
\lambda_3(t_3-t)]^{\frac{1}{D(D-2)}}}{[\sinh q\cosh\theta
\lambda_1(t_1-t)]^{\frac{(D-1)^2}{D(D-2)(1-g^2)}}}\,.
\end{equation}
Assuming $t_1<t_3$, the scale factor $e^{\tilde{a}}$ increases monotonically
in $-\infty<t<t_1$.
In terms of the cosmic time $\tau$ ($d\tau=\pm e^{(D-1)\tilde{a}}dt$),
the scale factor has behavior  $e^{\tilde{a}}\sim \tau^{1/(D-1)}$ for $\tau\sim 0$,
while $e^{\tilde{a}}\sim \tau^{(D-1)^2/(1+D(D-2)g^2)}$ for $\tau\rightarrow\infty$.
Because $(D-1)^2/(1+D(D-2)g^2)>1$, we find that the solution describes power-law
inflation. Unfortunately, non-zero coupling $g$ decreases the power.
Note that the case with $g=0$ and $\phi\equiv 0$ reduces the model into the one of 
pure $R^p$ gravity, and the effective potential for $\psi$ is very similar to that
studied by Mignemi and Pintus \cite{MP,PM}.

%%%%%%%%%%%%%%%%%%%%%%%%%%%%%%%%%%%%%%%%%%%%%%%%%%%%%%%%%%%%%%%%%%%%%%%%%%%

Now, we turn to quantum cosmology in the present case.
The WDW equation is obtained by $\hat{H}\Psi=0$, where
$\hat{H}$ is the Hamiltonian (\ref{hami}) in which momenta are replaced by
differential operators.
Owing to the separation of variables, the general solution for the WDW equation
can be obtained easily as
\begin{eqnarray}
&
&\Psi(x,y,z)=\int\int[A(q,\Theta)J_{iq\cosh\Theta/\lambda_1}(e^{\lambda_1x}/\lambda_1)
+B(q,\Theta)J_{-iq\cosh\Theta/\lambda_1}(e^{\lambda_1x}/\lambda_1)]\nonumber \\ &
&\times[C(q,\Theta)J_{iq\sinh\Theta/\lambda_3}(e^{\lambda_3z}/(\sqrt{D-1/2}\lambda_3))
+D(q,\Theta)J_{-iq\sinh\Theta/\lambda_3}(e^{\lambda_3z}/(\sqrt{D-1/2}\lambda_3))]
\nonumber \\
& &
\times e^{iqy}dq d\Theta\,,
\end{eqnarray}
where $A$, $B$, $C$, and $D$ are amplitudes.
This form of solution corresponds to the case ($E_1<0, E_3>0$).
Each wave mode is oscillatory, because there is no potential `wall' at all.

%%%%%%%%%%%%%%%%%%%%%%%%%%%%%%%%%%%%%%%%%%%%%%%%%%%%%%%%%%%%%%%%%%%%%%%%%%%
%%%%%%%%%%%%%%%%%%%%%%%%%%%%%%%%%%%%%%%%%%%%%%%%%%%%%%%%%%%%%%%%%%%%%%%%%%%
%%%%%%%%%%%%%%%%%%%%%%%%%%%%%%%%%%%%%%%%%%%%%%%%%%%%%%%%%%%%%%%%%%%%%%%%%%%
\section{Summary and discussion}
\label{dis}
%%%%%%%%%%%%%%%%%%%%%%%%%%%%%%%%%%%%%%%%%%%%%%%%%%%%%%%%%%%%%%%%%%%%%%%%%%%
%%%%%%%%%%%%%%%%%%%%%%%%%%%%%%%%%%%%%%%%%%%%%%%%%%%%%%%%%%%%%%%%%%%%%%%%%%%
%%%%%%%%%%%%%%%%%%%%%%%%%%%%%%%%%%%%%%%%%%%%%%%%%%%%%%%%%%%%%%%%%%%%%%%%%%%

We have shown that a modification of higher order terms in $R$
can bring about interesting cosmological models.
The $K$-essential modification, which utilizes a kinetic-term-like combination
of a scalar field, realizes a simple addition of a canonical scalar field into
the theory. In this paper, we have studied various classical and quantum
cosmologies derived from the actions which contain a $K$-essentially modified
term of scalar curvature squared or specific powered.
Even though the Starobinsky model can explain the recent observations
very well, to introduce modifications in the model is a good way to study its
robustness and special properties.

%R1
%The simple kinetic term of the additional scalar field can cause
%separability of two scalar dynamics especially in the case with $\gamma=2$, where
%$U_{\psi}(0,\phi)=0$.
%Provided that the scalaron rapidly loses energy by other
%matter fields as usually expected and reaches an expectation value
%$\langle\psi\rangle=0$, the equation of motion of the scalar field $\phi$ is
%still $\ddot{\phi}+3H\dot{\phi}+V''(\phi)\approx 0$ (where $H$ is the expansion
%rate). Although some hierarchical settings (e.g., $\beta^{-1}\gg V''$) is required
%in this case, there is a possibility of identifying the scalar field with a
%quintessence \cite{darkenergy1,darkenergy2}. Thus, it can be said that our model
%cannot explain the hierarchy of two acceleration phase but can `permit' the
%hierarchy.
%R1

An interesting outcome in the present work is finding that there is a class of
solvable models in higher order theory with an additional scalar mode.
This is due to the fact that the scalar fields have canonical kinetic terms.
This simplicity can be a strong motivation to study
the solutions for compact objects with strong gravity, such as black holes, 
in our models and their extensions.

As subjects for future research, we can also consider the following 
generalization: the
$K$-essential modification of supergravity extensions \cite{FKR,FKP1,FKP2}
of higher order theory,
higher dimensional models with and without dimensional reduction \cite{KN,PPW},
higher derivative correction \cite{CNMOO,CSSS,CMP},
higher order theories with other than scalar curvature (e.g. \cite{ost}).

%%%%%%%%%%%%%%%%%%%%%%%%%%%%%%%%%%%%%%%%%%%%%%%%%%%%%%%%%%%%%%%%%
%%%%%%%%%%%%%%%%%%%%%%%%%%%%%%%%%%%%%%%%%%%%%%%%%%%%%%%%%%%%%%%%%
\appendix
%%%%%%%%%%%%%%%%%%%%%%%%%%%%%%%%%%%%%%%%%%%%%%%%%%%%%%%%%%%%%%%%%
%%%%%%%%%%%%%%%%%%%%%%%%%%%%%%%%%%%%%%%%%%%%%%%%%%%%%%%%%%%%%%%%%

\section{Solvable pure $R^2$ gravity in different variables and 
the factor ordering}
\label{AA}

A different factor ordering gives a different solution of the WDW
equation, at least in a certain region of variables.
In this Appendix \ref{AA}, we review the solvable model
of pure $R^2$ gravity in four dimensions as an example, and considered the factor
ordering when the different variables are used. 

For the flat four-dimensional spacetime, the pure $R^2$ gravity, i.e., in absence
of the additional scalar field, is known to be solvable in the minisuperspace
formalism
\cite{Schmidt1,Schmidt2}.
In such a case, the effective Lagrangian becomes
\begin{equation}
L'(a,\dot{a},Q,\dot{Q})=-2\left(\frac{Q}{a}\right)^{\cdot}\dot{a}
-\frac{1}{36\pi^2\beta}Q^2\,,
\end{equation}
since the term which comes from the spatial curvature is also absent.
If we use new variables $\sigma\equiv a^3$ and $\tau\equiv Q/a$,
the Lagrangian can be written as
\begin{equation}
L'(\sigma,\dot{\sigma},\tau,\dot{\tau})=-\frac{2}{3\sigma^{2/3}}\dot{\sigma}\dot{\tau}
-\frac{\sigma^{2/3}\tau^2}{36\pi^2\beta}\,,
\end{equation}
and the Hamiltonian is given by
\begin{equation}
H(\sigma,\tau,P_\sigma,P_\tau)=-\frac{3\sigma^{2/3}}{2}\left(P_{\sigma}P_{\tau}
-\frac{\tau^2}{54\pi^2\beta}\right)\,.
\end{equation}
Therefore, the WDW equation reads
\begin{equation}
\left(\frac{\partial}{\partial\sigma}\frac{\partial}{\partial\tau}+
\frac{\tau^2}{54\pi^2\beta}\right)\Psi(\sigma,\tau)=0\,.
\end{equation}
This equation can be exactly in the form
\begin{equation}
\Psi(\sigma,\tau)=\int_{-\infty}^\infty
A(\lambda)\exp\left[-\lambda\sigma+\frac{\tau^3}{162\pi^2\beta\lambda}
\right]d\lambda\,,
\end{equation}
where $A(\lambda)$ indicates the amplitude \cite{Schmidt1,Schmidt2}.

Here, we consider the Fourier transform of the elementary wave solution.
That is
\begin{equation}
w=\int_{-\infty}^\infty
\exp\left[-\lambda\sigma+\frac{\tau^3}{162\pi^2\beta\lambda}
\right]e^{-il\theta}d\theta\,.
\end{equation}
The definitions of the variables are the same as in the previous sections, i.e.,
\begin{equation}
x=\frac{Q}{a}-a=r\sinh\theta\,,\quad
y=a+\frac{Q}{a}=r\cosh\theta\,.
\label{ct}
\end{equation}
Thus, we find
\begin{equation}
\sigma=a^3=\frac{r^3}{8}e^{-3\theta}\,,\quad\tau=\frac{Q}{a}=\frac{r}{2}e^{\theta}
\,.
\end{equation}
Then, we obtain
\begin{eqnarray}
w&=&\int_{-\infty}^\infty
\exp\left[-\lambda\frac{r^3}{8}e^{-3\theta}
+\frac{r^3e^{3\theta}}{8\cdot 162\pi^2\beta\lambda}
-il\theta\right]d\theta
\nonumber \\
%&=&\int_{-\infty}^\infty
%\exp\left[-il\theta-i\frac{r^3}{36\sqrt{2\beta}\pi}\frac{1}{2}
%\left(-i9\sqrt{2\beta}\pi\lambda
%e^{-3\theta} +\frac{e^{3\theta}}{-i9\sqrt{2\beta}\pi\lambda}\right)
%\right]d\theta
%\nonumber \\
&=&\frac{1}{3}(-i9\sqrt{2\beta}\pi\lambda)^{-il/3}\int_{-\infty}^\infty
\exp\left[-i\frac{l}{3}t-i\frac{r^3}{36\sqrt{2\beta}\pi}\cosh t
\right]dt
\nonumber \\
&\propto& K_{il/3}(ir^3/(36\sqrt{2\beta}\pi))\,.
\end{eqnarray}
Because $K_{il/3}(ir^3/(36\sqrt{2\beta}\pi))$ is expressed by
a linear combination of $J_{\pm il/3}(r^3/(36\sqrt{2\beta}\pi))$,
general solution can be written by
\begin{eqnarray}
\Psi&=&\int\int \left[
A(l,m) J_{i\frac{l}{3}}(r^3/(36\sqrt{2\beta}\pi))\right.
\nonumber \\
& &\qquad\quad\left.+
B(l,m) J_{-i\frac{l}{3}}(r^3/(36\sqrt{2\beta}\pi))\right]
e^{il\theta} dl\,.
\end{eqnarray}
This is the form of the general explicit solution of the following equation:
\begin{equation}
\left[\frac{1}{r}\frac{\partial}{\partial
r}r\frac{\partial}{\partial
r}-\frac{1}{r^2}\frac{\partial^2}{\partial\theta^2}+\frac{r^4}{288\pi^2\beta}
\right]\Psi(r,\theta)=0\,.
\label{ge}
\end{equation}

We conclude that the exact solution obtained by Schmidt \cite{Schmidt1,Schmidt2}
is equivalent of the solution of (\ref{ge}), where the parameter of ordering $s$
equals to one,%
\footnote{Indeed, we find that
$\frac{\partial^2}{\partial x^2}-\frac{\partial^2}{\partial y^2}=
\frac{1}{r}\frac{\partial}{\partial r}r\frac{\partial}{\partial
r}-\frac{1}{r^2}\frac{\partial^2}{\partial \theta^2}$ by the coordinate
transformation (\ref{ct}).}
 whereas the authors of \cite{Vilenkin,MMS2} chose the different
factor ordering ($s=0$) in the `kinetic' term (in the Starobinsky model).

%%%%%%%%%%%%%%%%%%%%%%%%%%%%%%%%%%%%%%%%%%%%%%%%%%%%%%%%%%%%%%%%%%%%%%%%%%%
%\acknowledgments
%%%%%%%%%%%%%%%%%%%%%%%%%%%%%%%%%%%%%%%%%%%%%%%%%%%%%%%%%%%%%%%%%%%%%%%%%%%
%Acknowledgements
%%%%%%%%%%%%%%%%%%%%%%%%%%%%%%%%%%%%%%%%%%%%%%%%%%%%%%%%%%%%%%%%%%%%%%%%%%%
%\begin{acknowledgments}
%We thank
%the organizers of JGRG21, where our
%partial result %({\tt [arXiv:10mm.xxxx]}) 
%was presented. %for elucidating comments.
%This study is supported in part by the Grant-in-Aid of Nikaido Research 
%Fund.
%\end{acknowledgments}
%%%%%%%%%%%%%%%%%%%%%%%%%%%%%%%%%%%%%%%%%%%%%%%%%%%%%%%%%%%%%%%%%%%%%%%%%%%

%%%%%%%%%%%%%%%%%%%%%%%%%%%%%%%%%%%%%%%%%
%%%%%%%%%%%%%%%%%%%%%%%%%%%%%%%%%%%%%%%%%
%%%
%%%   References
%%%
%%%%%%%%%%%%%%%%%%%%%%%%%%%%%%%%%%%%%%%%%
%%%%%%%%%%%%%%%%%%%%%%%%%%%%%%%%%%%%%%%%%
%%%%%%%%%%%%%%%%%%%%%%%%%%%%%%%%%%%%%%%%%%%%%%%%%%%%%%%%%%%%%%%%%%%%%%%%%%%
%thebibliography
%%%%%%%%%%%%%%%%%%%%%%%%%%%%%%%%%%%%%%%%%%%%%%%%%%%%%%%%%%%%%%%%%%%%%%%%%%%
%\bibliographystyle{apsrev}
\bibliographystyle{apsrev4-1}
%\bibliography{}

\begin{thebibliography}{99}

%%%%%%%%%%%%%%%%%%%%%%%%%%%%%%%%%%%%%%%%%%%%%%%%%%%%%%%%%%%%%%%%%%%%%

\bibitem{inflation} A.~D.~Linde,
``Particle physics and inflationary cosmology'',
Contemporary Concepts in Physics \textbf{5}, % (1990) 1.%--362.
(Harwood Academic Pub., Philadelphia, 1990).
%arXiv:hep-th/0503203.

\bibitem{EW} J.~Ellis and D.~Wands,
``22. Inflation'',
in \textit{Review of particle physics}, 
M.~Tanabashi et al.~(Particle Data Group), Phys. Rev. \textbf{D98} (2018) 030001,
pp.~364--376.

%%%%%%%%%%%%%%%%%%%%%%%%%%%%%%%%%%%%%%%%%%%%%%%%%%%%%%%%%%%%%%%%%%%%%

\bibitem{darkenergy1} L.~Amendra and S.~Tsujikawa,
``Dark energy: theory and observations'',
(Cambridge University Press, New York, 2010).

\bibitem{darkenergy2} E.~J.~Copeland, M.~Sami and S.~Tsujikawa,
%``Dynamics of dark energy'',
Int. J. Mod. Phys. \textbf{D15} (2006) 1753. %--1935.
%DOI: 10.1142/S021827180600942X
%arXiv:hep-th/0603057.

\bibitem{darkenergy3} D.~H.~Weinberg and M.~White,
``27. Dark Energy'',
in \textit{Review of particle physics}, 
M.~Tanabashi et al.~(Particle Data Group), Phys. Rev. \textbf{D98} (2018) 030001,
pp.~406--413.

%%%%%%%%%%%%%%%%%%%%%%%%%%%%%%%%%%%%%%%%%%%%%%%%%%%%%%%%%%%%%%%%%%%%%
%NO1,SF,CLF,FT,NO2,CL,CFPS,Koyama,NOO,Ishak
\bibitem{NO1}
S.~Nojiri and S.~D.~Odintsov,
%Introduction to modified gravity and gravitational
%alternative for dark energy
Int. J. Geom. Methods Mod. Phys. {\bf 04} (2007) 115.
%arXiv:hep-th/0601213.

\bibitem{SF}
T.~P.~Sotiriou and V.~Faraoni,
%f(R) Theories Of Gravity
Rev. Mod. Phys. {\bf 82} (2010) 451.
%arXiv:0805.1726 [gr-qc].

\bibitem{CLF}
S.~Capozziello, M.~De Laurentis and V.~Faraoni,
%%A bird's eye view of $f(R)$ gravity.
Open Astron. J. {\bf 3} (2010) 49. %--72.
%arXiv:0909.4672 [gr-qc].

\bibitem{FT}
A.~De Felice and S.~Tsujikawa,
%f(R) theories
Living Rev. Rel. {\bf 13} (2010) 3.
%arXiv:1002.4928 [gr-qc]

\bibitem{NO2}
S.~Nojiri and S.~D.~Odintsov,
%Unified cosmic history in modified gravity: 
%from F(R) theory to Lorentz non-invariant models
Phys. Rep. {\bf 505} (2011) 59.
%arXiv:1011.0544v4 [gr-qc].

\bibitem{CL}
S.~Capozziello and M.~De Laurentis,
%%Extended Theories of Gravity.
Phys. Rep. {\bf 509} (2011) 167.
%arXiv:1108.6266 [gr-qc].

\bibitem{CFPS}
T.~Clifton, P.~G.~Ferreir, A.~Padilla and C.~Skordis,
%%Modified gravity and cosmology
Phys. Rep. {\bf 515} (2012) 1.
%arXiv:1106.2476 [astro-ph].

\bibitem{Koyama}
K.~Koyama,
%Cosmological tests of modified gravity
Rept. Prog. Phys. \textbf{79} (2016) 046902.
%arXiv:1504.04623 [astro-ph.CO].

\bibitem{NOO}
S.~Nojiri, S.~D.~Odintsov and V.~K.~Oikonomou,
%Modified gravity theories on a nutshell:
%Inflation, bounce and late-time evolution
Phys. Rep. {\bf 692} (2017) 1.
%arXiv:1705.11098 [gr-qc].

\bibitem{Ishak}
M.~Ishak,
%``Testing general relativity in cosmology'',
Living Rev. Rel. \textbf{22} (2019) 1.
%arXiv:1806.10122 [astro-ph.CO].

%%%%%%%%%%%%%%%%%%%%%%%%%%%%%%%%%%%%%%%%%%%%%%%%%%%%%%%%%%%%%%%%%%%%%%%%%%%

\bibitem{Starobinsky}
A.~A.~Starobinsky,
%``A new type of isotropic cosmological models without singularity'',
Phys. Lett. \textbf{B91} (1980) 99. %--102.

%%%%%%%%%%%%%%%%%%%%%%%%%%%%%%%%%%%%%%%%%%%%%%%%%%%%%%%%%%%%%%%%%%%%%%%%%%%

\bibitem{Vilenkin}
A.~Vilenkin,
%``Classical and quantum cosmology of the Starobinsky model'',
Phys. Rev. \textbf{D32} (1985) 2511. %--2521.

\bibitem{MMS1}
M.~B.~Miji\'c, M.~S.~Morris and W.-M.~Suen,
%``The $R^2$ cosmology: Inflation without a phase transition'',
Phys. Rev. \textbf{D34} (1986) 2934. %--2946.

%%%%%%%%%%%%%%%%%%%%%%%%%%%%%%%%%%%%%%%%%%%%%%%%%%%%%%%%%%%%%%%%%%%%%%%%%%%

\bibitem{BP}
C.~van de Bruck and L.~E.~Paduraru,
%``Simplest extension of Starobinsky inflation'',
Phys. Rev. \textbf{D92} (2015) 083513.
%arXiv:1505.01727 [hep-th].

\bibitem{BDP}
C.~van de Bruck, P.~Dunsby and L.~E.~Paduraru,
%``Reheating and preheating in the simplest extension of Starobinsky inflation'',
Int. J. Mod. Phys. \textbf{D26} (2017) 1750152.
%arXiv:1606.04346 [gr-qc].

\bibitem{MKW}
T.~Mori, K.~Kohri and J.~White,
%``Multi-field effects in a simple extension of $R^2$ inflation'',
JCAP \textbf{1710} (2017) 044.
%arXiv:1705.05638 [astro-ph.CO].

\bibitem{KK}
S.~Kaneda and S.~V.~Ketov,
%``Starobinsky-like two-field inflation'',
Eur. Phys. J. \textbf{C76} (2016) 26.
%arXiv:1510.03524 [hep-th].

%%%%%%%%%%%%%%%%%%%%%%%%%%%%%%%%%%%%%%%%%%%%%%%%%%%%%%%%%%%%%%%%%%%%%%%%%%%
%CCH,MMS,SN
\bibitem{CCH}
V.~H.~C\'ardenas, S.~del Campo and R.~Herrera,
%``$R^2$-corrections to chaotic inflation'',
Mod. Phys. Lett. \textbf{A18} (2003) 2039. %--2050.
%arXiv:gr-qc/0308040.

\bibitem{MMS}
S.~Myrzakul, R.~Myrzakulov and L.~Sebastiani,
%``Chaotic inflation in higher derivative gravity theories'',
Eur. Phys. J. \textbf{C75} (2015) 111.
%arXiv:1501.01796 [gr-qc].

\bibitem{SN}
M.~Sharif and I.~Nawazish,
%``The view of chaotic inflationary universe from $f(R)$ gravity'', 
Astrophys. Space Sci. \textbf{361} (2016) 19. 

%%%%%%%%%%%%%%%%%%%%%%%%%%%%%%%%%%%%%%%%%%%%%%%%%%%%%%%%%%%%%%%%%%%%%%%%%%%
%GT1,BOT,SM,Wang,Ema,Salvio,MSY,GT2,GS
\bibitem{GT1}
D.~Gorbunov and A.~Tokareva,
%``Scale-invariance as the origin of dark radiation?'',
Phys. Lett. \textbf{B739} (2014) 50. %--55.
%arXiv:1307.5298 [astro-ph.CO].

\bibitem{BOT}
K.~Bamba, S.~D.~Odintsov and P.~V.~Tretyakov,
%``Inflation in a conformally invariant two-scalar-field theory with an extra
%$R^2$ term'',
Eur. Phys. J. \textbf{C75} (2015) 344.
%arXiv:1505.00854 [hep-th].

\bibitem{SM}
A.~Salvio and A.~Mazumdar,
%``Classical and quantum initial conditions for Higgs inflation'',
Phys. Lett. \textbf{B750} (2015) 194. %--200. 
%arXiv:1506.07520 [hep-ph].

\bibitem{Wang}
Y.-C. Wang and T.~Wang,
%``Primordial perturbations generated by Higgs field and $R^2$ operator'',
Phys. Rev. \textbf{D96} (2017) 123506. %(12 pages)
%arXiv:1701.06636 [gr-qc].

\bibitem{Ema}
Y.~Ema,
%``Higgs scalaron mixed inflation'',
Phys. Lett. \textbf{B770} (2017) 403. %--411.
%arXiv:1701.07665 [hep-ph].

\bibitem{Salvio}
A.~Salvio,
%``Initial conditions for critical Higgs inflation'',
Phys. Lett. \textbf{B780} (2018) 111. %--117. 
%arXiv:1712.04477 [hep-ph].

\bibitem{MSY}
M.~He, A.~A.~Starobinsky and J.~Yokoyama,
%``Inflation in the mixed Higgs-$R^2$ model'',
JCAP \textbf{1805} (2018) 064.
%arXiv:1804.00409 [astro-ph.CO].

\bibitem{GT2}
D.~Gorbunov and A.~Tokareva,
%``Scalaron the healer: removing the strong-coupling in the Higgs- and
%Higgs-dilaton inflations'',
Phys. Lett. \textbf{B788} (2019) 37. %--41. 
%arXiv:1807.02392 [hep-ph].

\bibitem{GS}
A.~Gundhi and C.~F.~Steinwachs,
%``Scalaron-Higgs inflation'',
arXiv:1810.10546 [hep-th].


%%%%%%%%%%%%%%%%%%%%%%%%%%%%%%%%%%%%%%%%%%%%%%%%%%%%%%%%%%%%%%%%%%%%%

\bibitem{CEOS}
\'A.~Cruz-Dombriz, E.~Elizalde, S.~D.~Odintsov and D.~S\'aez-G\'omez,
%``Spotting deviation from $R^2$ inflation'',
JCAP \textbf{1605} (2016) 060.
%arXiv:1603.05537 [gr-qc].

\bibitem{AMS}
C.~Armendariz-Picon, V.~Mukhanov and P.~J.~Steinhardt,
%``Dynamical solution to the problem of a small cosmological constant and
%late-time cosmic acceleration'', 
Phys. Rev. Lett. \textbf{85} (2000) 4438. %--4441.

%%%%%%%%%%%%%%%%%%%%%%%%%%%%%%%%%%%%%%%%%%%%%%%%%%%%%%%%%%%%%%%%%%%%%
%GW,STY,WBMR,MF,LLPT,Wands,BR,WW
\bibitem{GW}
J.~Garc\'{\i}a-Bellido and D.~Wands,
%``Metric perturbations in two-field inflation'',  
Phys. Rev. \textbf{D53} (1996) 5437. %--5445.
%arXiv:astro-ph/9511029.

\bibitem{STY}
A.~A.~Starobinsky, S.~Tsujikawa and J.~Yokoyama,
%``Cosmological perturbations from multi-field inflation in generalized Einstein
%theories'',  
Nucl. Phys. \textbf{B610} (2001) 383. %--410.
%arXiv:astro-ph/0107555.

\bibitem{WBMR}
D.~Wands, N.~Bartolo, S.~Matarrese and A.~Riotto,
%``Observational test of two-field inflation'',  
Phys. Rev. \textbf{D66} (2002) 043520.
%arXiv:astro-ph/0205253.

\bibitem{MF}
F.~Di Marco and F.~Finelli,
%``Slow-roll inflation for generalized two-field Lagrangians'',  
Phys. Rev. \textbf{D71} (2005) 123502.
%arXiv:astro-ph/0505198.

\bibitem{LLPT}
Z.~Lalak, D.~Langlois, S.~Pokorsky and K.~Turzy\'nski,
%``Curvature and isocurvature perturbations in two-field inflation'', 
JCAP \textbf{0707} (2007) 014.
%arXiv:0704.0212 [hep-th].

\bibitem{Wands}
T.~Wang,
%``Note on non-Gaussianities in two-field inflation'',  
Phys. Rev. \textbf{D82} (2010) 123515.
%arXiv:1008.3198 [astro-ph].

\bibitem{BR}
C.~van de Bruck and M.~Robinson,
%``Power spectra beyond the slow roll approximation in theories with
%non-canonical kinetic terms'', 
JCAP \textbf{1408} (2014) 024.
%arXiv:1404.7806 [astro-ph.CO].

\bibitem{WW}
Y.-C.~Wang and T.~Wang,
%``Non-canonical two-field inflation to order $\xi^2$'', 
Int. J. Mod. Phys. \textbf{D27} (2018) 1850026.
%arXiv:1603.0956 [gr-qc].

%%%%%%%%%%%%%%%%%%%%%%%%%%%%%%%%%%%%%%%%%%%%%%%%%%%%%%%%%%%%%%%%%%%%%

\bibitem{HL}
S.~W.~Hawking and J.~C.~Luttrell,
%``Higher derivatives in quantum cosmology'',
Nucl. Phys. \textbf{B247} (1984) 250. %--260.

\bibitem{HW}
S.~W.~Hawking and Z.~C.~Wu,
%``Numerical calculations of minisuperspace cosmological models'',
Phys. Lett. \textbf{B151} (1985) 15. %--20.

\bibitem{MMS2}
M.~B.~Miji\'c, M.~S.~Morris and W.-M.~Suen,
%``Initial conditions for $R+\epsilon R^2$ cosmology'', 
Phys. Rev. \textbf{D39} (1989) 1496. %--1510.

%%%%%%%%%%%%%%%%%%%%%%%%%%%%%%%%%%%%%%%%%%%%%%%%%%%%%%%%%%%%%%%%%%%%%

\bibitem{York}
J.~W.~York,
%``Role of conformal three-geometry in the dynamics of gravitation'',
Phys. Rev. Lett. \textbf{28} (1972) 1082.

\bibitem{GH}
G.~W.~Gibbons and S.~W.~Hawking,
%``Action integrals and partition functions in quantum gravity'',
Phys. Rev. \textbf{D15} (1977) 2752.

%%%%%%%%%%%%%%%%%%%%%%%%%%%%%%%%%%%%%%%%%%%%%%%%%%%%%%%%%%%%%%%%%%%%%%%%%%%

\bibitem{HKY}
D.-I.~Hwang, S.~A.~Kim and D.-H.~Yeom,
%``No boundary wave function for two-field inflation'', 
Class. Quant. Grav. \textbf{32} (2015) 115006. 
%arXiv:1404.2800 [gr-qc].

%%%%%%%%%%%%%%%%%%%%%%%%%%%%%%%%%%%%%%%%%%%%%%%%%%%%%%%%%%%%%%%%%%%%%%%%%%%

\bibitem{Kiefer}
C.~Kiefer,
%``Wave packets in quantum cosmology and the cosmological constant'',
Nucl. Phys. \textbf{B341} (1990) 273. %--293.

%%%%%%%%%%%%%%%%%%%%%%%%%%%%%%%%%%%%%%%%%%%%%%%%%%%%%%%%%%%%%%%%%%%%%

\bibitem{Horowitz}
G.~T.~Horowitz,
%``Quantum cosmology with a positive-definite action'',
Phys. Rev. \textbf{D31} (1985) 1169. %--1177.

\bibitem{Schmidt1}
H.-J.~Schmidt,
%``Stability and Hamiltonian formulation of higher derivative theories'',
Phys. Rev. \textbf{D49} (1994) 6354; %--63669.
%arXiv:gr-qc/9404038.

\bibitem{Schmidt2}
H.-J.~Schmidt,
%``Stability and Hamiltonian formulation of higher derivative theories'',
Phys. Rev. \textbf{D54} (1996) 7906(E).
%arXiv:gr-qc/9404038.

%%%%%%%%%%%%%%%%%%%%%%%%%%%%%%%%%%%%%%%%%%%%%%%%%%%%%%%%%%%%%%%%

\bibitem{KNw} Y.~Kazama and R.~Nakayama, 
%``Wave packet in quantum cosmology'',
Phys. Rev. \textbf{D32} (1985) 2500.

\bibitem{Kiefer1} C.~Kiefer, 
%``Wave packets in minisuperspace'',
Phys. Rev. \textbf{D38} (1988) 1761. %--1771.

%%%%%%%%%%%%%%%%%%%%%%%%%%%%%%%%%%%%%%%%%%%%%%%%%%%%%%%%%%%%%%%%%%%%%
%MP,PM,TW,Ohta2,LMPX,Kaloper,CGG,Roy,ANL,ALNW,KKST
\bibitem{MP}
S.~Mignemi and N.~Pintus,
%``An exactly solvable inflationary model'',
Gen. Rel. Grav. \textbf{47} (2015) 51. 
%DOI: 10.1007/s10714-015-1892-6 
%arXiv:1404.4720 [gr-qc].

\bibitem{PM}
N.~Pintus and S.~Mignemi,
%``Mathematical aspects of an exactly solvable inflationary model'',
Journal of Physics: Conf. Series \textbf{956} (2018) 012022.
%DOI: 10.1088/1742-6596/956/1/012022

%%%%%%%%%%%%%%%%%%%%%%%%%%%%%%%%%%%%%%%%%%%%%%%%%%%%%%%%%%%%%%%%

\bibitem{TW}
P.~K.~Townsend and M.~N.~R.~Wohlfarth, 
%``Accelerating cosmologies from compactification'',
Phys. Rev. Lett. \textbf{91} (2003) 061302.
%DOI: 10.1103/PhysRevLett.91.061302 
%hep-th/0303097.

\bibitem{Ohta2}
N.~Ohta,
%``Accelerating cosmologies from S-branes'',
Phys. Rev. Lett. \textbf{91} (2003) 061303. 
%DOI: 10.1103/PhysRevLett.91.061303 
%hep-th/0303238.

\bibitem{LMPX}
H.~L\"u, S.~Mukherji, C.~N.~Pope and K.~W.~Xu,
%``Cosmological solutions in string theories'',
Phys. Rev. \textbf{D55} (1997) 7926. %--7935.
%DOI: 10.1103/PhysRevD.55.7926 
%hep-th/9610107.

\bibitem{Kaloper}
N.~Kaloper,
%``Stringy Toda cosmologies'',
Phys. Rev. \textbf{D55} (1997) 3394. %--3402. 
%DOI: 10.1103/PhysRevD.55.3394 
%hep-th/9609087.

\bibitem{CGG}
C.-M.~Chen, D.~V.~Gal'tsov and M.~Gutperle,
%``S brane solutions in supergravity theories'', 
Phys. Rev. \textbf{D66} (2002) 024043.
%DOI: 10.1103/PhysRevD.66.024043 
%hep-th/0204071.

\bibitem{Roy}
S.~Roy,
%``Accelerating cosmologies from M/string theory compactifications'',
Phys. Lett. \textbf{B567} (2003) 322. %--329. 
%DOI: 10.1016/j.physletb.2003.06.060 
%hep-th/0304084.

%%%%%%%%%%%%%%%%%%%%%%%%%%%%%%%%%%%%%%%%%%%%%%%%%%%%%%%%%%%%%%%%

\bibitem{ANL}
A.~A.~Andrianov, O.~O.~Novikov and C.~Lan,
%`Quantum cosmology of multifield scalar matter: Some exact solutions'', 
Theor. Math. Phys. \textbf{184} (2015) 1224. %--1233, 
%Teor. Mat. Fiz. 184 (2015) no.3, 380-391. 
%DOI: 10.1007/s11232-015-0328-5 
%arXiv:1503.05527 [hep-th].

\bibitem{ALNW}
A.~A.~Andrianov, C.~Lan, O.~O.~Novikov and Y.-F.~Wang,
%``Integrable Cosmological Models with Field: Energy Density
%Self-Adjointness and Semiclassical Wave Packets'', 
Eur. J. Phys \textbf{C78} (2018) 786.
%arXiv:1802.06720 [hep-th].

%%%%%%%%%%%%%%%%%%%%%%%%%%%%%%%%%%%%%%%%%%%%%%%%%%%%%%%%%%%%%%%%

\bibitem{KKST}
N.~Kan, M.~Kuniyasu, K.~Shiraishi and K.~Takimoto,
%``Integrable higher-dimensional cosmology with separable variables in an
%Einstein-dilaton-antisymmetric field theory'', 
Phys. Rev. \textbf{D98} (2018) 044054.
%arXiv:1806.10263 [hep-th].

%%%%%%%%%%%%%%%%%%%%%%%%%%%%%%%%%%%%%%%%%%%%%%%%%%%%%%%%%%%%%%%%%%%%%

\bibitem{LM}
F.~Lucchin and S.~Matarrese,
%``Power-law inflation'',
Phys. Rev. \textbf{D32} (1985) 1316.

%%%%%%%%%%%%%%%%%%%%%%%%%%%%%%%%%%%%%%%%%%%%%%%%%%%%%%%%%%%%%%%%%%%%%
%FKR,FKP1,FKP2

\bibitem{FKR}
F.~Farakos, A.~Kehagias and A.~Riotto,
%``On the Starobinsky model of inflation from supergravity'',
Nucl. Phys. \textbf{B876} (2013) 187. %-200.
%arXiv:1307.1147 [hep-th].

\bibitem{FKP1}
S.~Ferrara, R.~Kallosh and A.~Van Proeyen,
%``On the supersymmetric completion of $R+R^2$ gravity and cosmology'',
JHEP \textbf{1311} (2013) 134.
%arXiv:1309.4052 [hep-th].

\bibitem{FKP2}
S.~Ferrara,  A.~Kehagias and M.~Porrati,
%``$R$ supergravity'',
JHEP \textbf{1508} (2015) 001.
%arXiv:1506.01566 [hep-th].

%%%%%%%%%%%%%%%%%%%%%%%%%%%%%%%%%%%%%%%%%%%%%%%%%%%%%%%%%%%%%%%%%%%%%

\bibitem{KN}
S.~V.~Ketov and H.~Nakada,
%``Inflation from $(R+\gamma R^n-2\Lambda)$ gravity in higher dimensions'', 
Phys. Rev. \textbf{D95} (2017) 103507.
%arXiv:1701.08239 [hep-th].

\bibitem{PPW}
S.~Paj\'on Otero,  F.~G.~Pedro and C.~Wieck,
%``$R+\alpha R^n$ inflation in higher-dimensional space-times'',
JHEP \textbf{1705} (2017) 058.
%arXiv:1702.08311 [hep-th].

%%%%%%%%%%%%%%%%%%%%%%%%%%%%%%%%%%%%%%%%%%%%%%%%%%%%%%%%%%%%%%%%
%,CNMOO,CSSS,CMP
%\bibitem{CMMP}
%R.~R.~Cuzinatto, C.~A.~M.~de Melo, L.~G.~Medeiros and
%P.~J.~Pompeia, 
%``$f(R,\nabla_{\mu_1}R,\dots,\nabla_{\mu_1}\dots\nabla_{\mu_n}R)$ 
%theories of gravity in Einstein frame'',  
%arXiv:1806.08850 [gr-qc].

\bibitem{CNMOO}
S.~V.~Chervon, A.~V.~Nikolaev, T.~I.~Mayorova, S.~D.~Odintsov and
V.~K.~Oikonomou, 
%``Kinetic scalar curvature extended $f(R)$ gravity'',  
Nucl. Phys. \textbf{B936} (2018) 597. %--614.
%arXiv:1810.01900 [gr-qc].

\bibitem{CSSS}
A.~R.~Romero Castellanos, F.~Sobreira, L.~Shapiro and A.~A.~Starobinsky,
%``On higher derivative corrections to the $R+R^2$ inflationary model'', 
JCAP \textbf{1812} (2018) 007.
%arXiv:1810.07787 [gr-qc].

\bibitem{CMP}
R.~R.~Cuzinatto, L.~G.~Medeiros and P.~J.~Pompeia,
%``Higher-order modified Starobinsky inflation'', 
JCAP \textbf{1902} (2019) 055.
%arXiv:1810.08911 [gr-qc].

%%%%%%%%%%%%%%%%%%%%%%%%%%%%%%%%%%%%%%%%%%%%%%%%%%%%%%%%%%%%%%%%%%%%%

\bibitem{ost}
N.~Kan and K.~Shiraishi,
%``An ostentatious model of cosmological scalar-tensor theory'',
Mod. Phys. Lett. \textbf{A34} (2019) 1950144. %(9 pages).
%arXiv:1807.10411 [qr-qc].
%DOI: 101142/S021773231950144X

%%%%%%%%%%%%%%%%%%%%%%%%%%%%%%%%%%%%%%%%%%%%%%%%%%%%%%%%%%%%%%%%%%%%%

\end{thebibliography}

%%%%%%%%%%%%%%%%%%%%%%%%%%%%%%%%%%%%%%%%%%%%%%%%%%%%%%%%%%%%%%%%%%%%%%%%%%%
%%%%%%%%%%%%%%%%%%%%%%%%%%%%%%%%%%%%%%%%%%%%%%%%%%%%%%%%%%%%%%%%%%%%%%%%%%%
%%%%%%%%%%%%%%%%%%%%%%%%%%%%%%%%%%%%%%%%%%%%%%%%%%%%%%%%%%%%%%%%%%%%%%%%%%%
\end{document}